\documentclass[preprint]{aastex61}
\usepackage{comment}
\usepackage{amsmath,amssymb}
\usepackage{graphicx}
\usepackage{textcomp}
\usepackage{booktabs}
\usepackage{bm}

\newcommand\aastex{AAS\TeX}

\newcommand\ltsima{$\; \buildrel <\over\sim \;$}
\newcommand\simlt{\lower.5ex\hbox{\ltsima}}
\newcommand\gtsima{$\; \buildrel >\over\sim \;$}
\newcommand\simgt{\lower.5ex\hbox{\gtsima}}

\shorttitle{\aastex\ BH kick}
\shortauthors{Koshimoto et al.}

\defcitealias{kos21}{K21}

\begin{document}

\title{Influence of Black Hole Kick Velocity on Microlensing Distributions}

\author[0000-0003-2302-9562]{Naoki Koshimoto}
\affiliation{Department of Earth and Space Science, Graduate School of Science, Osaka University, Toyonaka, Osaka 560-0043, Japan}

\author[0000-0001-8181-7511]{Norita Kawanaka}
\affiliation{National Astronomical Observatory of
Japan (NAOJ), 2-21-1, Osawa, Mitaka, Tokyo 181-8588,
Japan}
\affiliation{Department of Physics, Graduate School of Science Tokyo Metropolitan University 1-1,
Minami-Osawa, Hachioji-shi, Tokyo 192-0397}
\affiliation{Center for Gravitational Physics and Quantum Information, Yukawa Institute for Theoretical Physics, Kyoto University, 606-8502, Kyoto, Japan}

\author[0000-0002-6347-3089]{Daichi Tsuna}
\affiliation{TAPIR, Mailcode 350-17, California Institute of Technology, Pasadena, CA 91125, USA}
\affiliation{Research Center for the Early Universe (RESCEU), Graduate School of Science, The University of Tokyo, 7-3-1 Hongo, Bunkyo-ku, Tokyo 113-0033, Japan}

\begin{abstract}
The natal kick velocity distribution for black holes (BHs) is unknown regardless of its importance for understanding the BH formation process.
Gravitational microlensing is a unique tool for studying the distribution of BHs in our Galaxy, and the first isolated stellar-mass BH event, OGLE-2011-BLG-0462/MOA-2011-BLG-191 (OB110462), was recently identified by astrometric microlensing.
This study investigates how the natal kick velocity for Galactic BHs affects the microlensing event rate distribution.
We consider a Maxwell distribution with various average kick velocities, as well as the consequent variation of the spatial distribution of BHs.
We find that the event rate for the BH lenses toward the Galactic bulge decreases as $v_{\rm avg}$ increases, mainly due to the scale height inflation.
We focus on the unique microlensing parameters measured for OB110462, with microlens parallax $\pi_{\rm E}$ larger than 0.06 for its long timescale of $t_{\rm E} > 200~$ days.
We calculate the expected number of BH events occurring with parameters similar to OB110462 during the OGLE-IV survey by \citet{mro17, mro19} and compare it with the actual number that occurred, at least one.
Our fiducial model predicts 0.26, 0.19, 0.095, 0.020, and $1.8 \times 10^{-3}$ events occurring for $v_{\rm avg} =$ 25 km/sec, 50 km/sec, 100 km/sec, 200 km/sec, and 400 km/sec, respectively, which suggests that the average kick velocity is likely to be $v_{\rm avg} \lesssim 100~{\rm km/sec}$.
The expected number smaller than unity even at maximum might indicate our luckiness of finding OB110462, which can be tested with future surveys by e.g. the Roman space telescope.
\end{abstract}


\section{Introduction} 
\label{sec:intro}
Black holes (BHs) are the remnants of massive stars after their gravitational collapse.  Considering the stellar evolution theory, initial mass function, and star formation rate in our Galaxy, one can estimate the number of Galactic stellar-mass BHs as $10^{8-9}$ \citep{1983bhwd.book.....S, 1992eocm.rept...29V, 1998ApJ...496..155S, 2017MNRAS.468.4000C,2018MNRAS.480.2704L}.  

Among Galactic BHs, those in binary systems with mass transfer from their companions can be detected as X-ray binaries, and several dozens of X-ray binaries are confirmed to contain BHs so far \citep{2016A&A...587A..61C}.  From the mass distribution of BHs in transient low mass X-ray binaries, it has been implied that compact objects with mass of $3-5M_{\odot}$ are absent \citep{oze10, 2011ApJ...741..103F}. If this `lower mass gap' \citep{1998ApJ...499..367B} is not an observational bias but intrinsic, it should have some hints about the process of BH formation after the core collapse of a massive star \citep{2012ApJ...749...91F}.  In other studies, the spatial distribution of BH X-ray binaries is investigated \citep{2020MNRAS.496L..22G,2021ApJ...921..131J}.  Assuming that BH X-ray binaries originate from the Galactic plane, their Galactic heights (i.e., the distances from the Galactic plane) reflect the strength of kicks on BHs at their births.  However, the Galactic BHs appearing in the analyses above are significantly biased to those contained in close binary systems, observed as bright X-ray binaries.  In order to draw conclusions about the BH formation processes in our Galaxy in a more general way, one should also investigate the statistical properties of BH binaries without mass transfer because of their large orbital separation (i.e., noninteracting BH binaries), and BHs without companions (i.e., isolated BHs).  As for noninteracting BH binaries, they can be detected via astrometric observations of their luminous companions' sinusoidal motions.  There are some studies that have estimated the number of noninteracting BH binaries detectable by the astrometric satellite {\it Gaia} \citep{2017IAUS..324...41K, 2017ApJ...850L..13B, 2017MNRAS.470.2611M, 2018ApJ...861...21Y, 2018MNRAS.481..930Y, 2018arXiv181009721K, 2019ApJ...886...68A, 2019ApJ...885..151S, 2020PASJ...72...45S, 2020ApJ...905..134W, 2022ApJ...931..107C, shikauchi+22, shikauchi+23}.  So far, three sources had been identified as noninteracting BH binaries from the {\it Gaia} data \citep{2023MNRAS.518.1057E, 2023ApJ...946...79T, 2023MNRAS.521.4323E, 2024arXiv240410486G}, and some candidates of BH binaries have been reported \citep{2022arXiv220700680A, 2023MNRAS.518.2991S}.

Another important subset of Galactic BHs is those without companions, i.e., isolated BHs.  Detection of isolated BHs is much more difficult than those in binaries.  One of the methods is to detect the emission from an isolated BH accreting surrounding medium \citep{1971SvA....15..377S, 1975A&A....44...59M, 1978ApJ...221..234G, 1979MNRAS.189..123C, 1985MNRAS.217...77M, 1993A&A...277..477C, 1998ApJ...495L..85F, 1998A&A...331..535P, 1999ApJ...523L...7A, 2002MNRAS.334..553A, 2003ApJ...596..437C, 2005MNRAS.360L..30M, 2005ApJ...628..873M, 2012MNRAS.427..589B, 2013MNRAS.430.1538F, 2017MNRAS.470.3332I, 2018MNRAS.475.1251M, tsu18, 2019MNRAS.488.2099T, 2021ApJ...922L..15K}, although no isolated accreting BH in our Galaxy has been unambiguously detected so far.

Another promising method to find isolated BHs is to detect the microlensing events.  When a massive compact object passes in front of a background star, the stellar light is deflected due to general relativistic effects, and an apparent magnification (photometric microlensing) and/or a positional shift (astrometric microlensing) of a background star will be temporarily observed \citep{1936Sci....84..506E, 1986ApJ...304....1P, 1995A&A...294..287H, 1995AJ....110.1427M, 1995ApJ...453...37W}.  Some previous studies propose to find Galactic BHs by the observations of microlensing events \citep{2000ApJ...535..928G, 2002ApJ...576L.131A, 2010A&A...523A..33S}.  There are four quantities that are involved in each microlensing event: the lens mass $M_{\rm L}$, the distance to the lens $D_{\rm L}$, the distance to the source $D_{\rm S}$, and the relative proper motion between the lens and the source $\mu_{\rm rel}$.  In the case of photometric microlensing, the duration of an event is described by the Einstein radius crossing time,
\begin{eqnarray}
t_{\rm E}=\frac{\theta_{\rm E}}{\mu_{\rm rel}},
\end{eqnarray}
where $\theta_{\rm E}$ is the angular Einstein radius given by
\begin{eqnarray}
    \theta_{\rm E}=\sqrt{\kappa M_{\rm L}\pi_{\rm rel}},
\end{eqnarray}
with $\kappa=8.144~{\rm mas}~M_{\odot}^{-1}$ and $\pi_{\rm rel}=1~{\rm au}~\left(D_{\rm L}^{-1}-D_{\rm S}^{-1} \right)$. In a long timescale event such as the one due to a BH lens, $t_{\rm E}$ and the microlens parallax given by $\pi_{\rm E}\equiv \pi_{\rm rel}/\theta_{\rm E}$ are measurable in a photometric microlensing event. Considering that in most cases the distance from a source is well-constrained (e.g., $D_{\rm S}\simeq 8~{\rm kpc}$ when the survey is toward the Galactic bulge), there is still a parameter degeneracy in an event, which makes it difficult to deduce the lens properties from photometric microlensing.  This degeneracy can be broken by the detection of astrometric microlensing \citep{ben02, 2016ApJ...830...41L, 2017ApJ...843..145K, 2017Sci...356.1046S}, which enables us to measure the mass of a lens object as well as its distance.  Recently, thanks to the combination of photometric and astrometric lensing observations, the mass of an isolated BH acting as a lens was measured for the first time \citep[OGLE-2011-BLG-0462/MOA-2011-BLG-191;][]{sah22, lam22, lam22S, mro22, lam23}.  Some studies have tried to constrain the origin or formation channel of this isolated BH with observed mass and velocity \citep{2022ApJ...930..159A, 2023ApJ...946L...2V}.  Even if astrometric microlensing events are not detected extensively, a number of detections of photometric microlensing events can give some constraints on the mass function, spatial distribution, and velocity distribution of Galactic BHs, which will provide insights into the physical processes occurring at the formation of BHs such as mass ejection, natal kick, and so on \citep{wyr16, lam20, 2020A&A...636A..20W, 2021AcA....71...89M, 2021arXiv210713697M, ros22, 2024ApJ...961..179P}.  Among them, the natal kick velocity distribution is of great importance for understanding the BH formation process.  It is expected that the compact remnants of massive stars (i.e., neutron stars and BHs) suffer from a kick at their births to have a peculiar velocity, possibly due to the asymmetry in the supernova explosion (for a review, see \citealp{2001ApJ...549.1111L}).  From the measurements of proper motions of young pulsars, the kick velocity of a neutron star is up to a few hundred ${\rm km}~{\rm sec}^{-1}$ \citep{2005MNRAS.360..974H}.  Although there are some works on the kick velocities of Galactic BHs based on the observations of X-ray binaries \citep{2005ApJ...625..324W, 2009ApJ...697.1057F, 2012ApJ...747..111W, 2014ApJ...790..119W, 2023ApJ...952L..34K, 2023ApJ...959..106B, 2023MNRAS.525.1498Z}, the kick velocity distribution of Galactic BHs is still uncertain both observationally and theoretically.  Very recently \cite{2024arXiv240314612S} predicted the number of observable microlensing events caused by Galactic isolated BHs or neutron stars making use of the population synthesis by \cite{2022MNRAS.516.4971S} which takes into account the spatial distributions of Galactic compact objects affected by natal kicks at their births.  Their results are obtained from a single kick velocity distribution model that is based on the observations of the neutron stars' proper motions \citep{2020MNRAS.494.3663I, 2021MNRAS.508.3345I}, and the assumption that natal kicks impart the same momentum to compact objects. 

In this study, we predict the parameter distributions of observable photometric microlensing events caused by Galactic isolated BHs. Especially, the difference in the kick velocity distributions gives rise to differences in the spatial distributions of BHs in the Galaxy, which affects the detectability of microlensing events caused by BHs as well as their durations and parallaxes.  We investigate the dependence of these observational properties on the kick velocity distribution.  In addition, we compare our results with past observations, including OGLE-2011-BLG-0462/MOA-2011-BLG-191, the microlensing event that is confirmed to be caused by a BH lens.

The structure of this paper is as follows.  In Section 2 we describe the Galactic model, the model of BH natal kicks, and the model of the spatial distribution of BHs employed in this study.  The influence of natal kicks on the parameter distributions of microlensing events is investigated in Section 3.  Section 4 is dedicated to the implication of our results on the natal kick velocity of the BH from the astrometric microlensing event OGLE-2011-BLG-0462/MOA-2011-BLG-191.  In Section 5 we discuss the uncertainties involved in the model, the possibility that OGLE-2011-BLG-0462/MOA-2011-BLG-191 is a `lucky' event, and comparisons with previous studies.  We summarize and conclude in Section 6.

\section{Model} \label{sec:model}
In this paper, we refer to a combination of the stellar mass function, stellar density distribution, and stellar velocity distribution in our Galaxy as a Galactic model.
We use the Galactic model developed by \citet{kos21} (hereafter \citetalias{kos21}) to simulate the microlensing events toward the Galactic bulge and we take their $E+E_{\rm X}$ model.
The model consists of a multi-component thin disk, a thick disk, and a barred bulge with an X-shaped structure.
It is optimized for use in microlensing studies toward the Galactic bulge region. It is fitted to various observational distributions toward the Galactic bulge, including OGLE-III red clump star count data \citep{nat13}, VIRAC proper motion data \citep{smi18, cla19}, BRAVA radial velocity data \citep{ric07, kun12}, and OGLE-IV star count and microlensing rate data \citep{mro17, mro19}.
It is particularly notable that this model is confirmed to reproduce the OGLE-IV $t_{\rm E}$ distribution \citep{mro17, mro19}, which enables us to constrain the kick velocity for BHs in Section \ref{sec:disc_vave}. 

\subsection{BH mass function} \label{sec:BHMFmodel}
The model by \citetalias{kos21} combines the stellar initial mass function, which has a power-law shape of $dN/dM_{\rm ini} \propto M_{\rm ini}^{-2.32}$ at $M_{\rm ini} > 0.86 M_{\odot}$, with the stellar age distribution and the initial-final mass relation (IFMR) by \citet{lam20} to obtain the present-day mass function for remnant objects.
For the age distribution, a time-dependent star formation rate $\propto \exp \left[ T/(7~{\rm Gyr}) \right]$ \citep{bov17} is applied to the thin disk, while a mono-age of 12~Gyr for the thick disk and a Gaussian distribution of $9 \pm 1~{\rm Gyr}$ for the bar are respectively assumed.
The IFMR is based on \citet{rai18} who combined observational data of BHs and neutron stars with 1-D neutrino-driven supernova simulations by \citet{suk16}.

In the IFMR, stars with initial masses of $15 < M_{\rm ini}/M_{\odot} < 120$ probabilistically evolve into BHs with masses between $\sim 5~M_{\odot}$ and $\sim 16~M_{\odot}$ (see Figure 6 of \citealt{lam20} for the IFMR for BHs).
The probabilities of being a BH are $P_{\rm evo, BH} = $ 0.321, 0.167, 0.500, 1.000, 0.348, 1.000, and 0.600 for initial mass ranges of 15 -- 17.8 $M_{\odot}$, 17.8 -- 18.5 $M_{\odot}$, 18.5 -- 21.7 $M_{\odot}$, 21.7 -- 25.2 $M_{\odot}$, 25.2 -- 27.5 $M_{\odot}$, 27.5 -- 60 $M_{\odot}$, and 60 -- 120 $M_{\odot}$, respectively.
The present-day mass function expects $\sim$0.0035 BHs per main sequence star in a typical line of sight toward the Galactic bulge.

\subsection{BH kick velocity}
In the original model by \citetalias{kos21}, a natal kick velocity of 100 km/sec in a random direction is added to the velocity of the BH progenitor following the prescription of \citet{lam20}.
In this paper, we consider five different values of the average kick velocity, $v_{\rm avg} =$ 25 km/sec, 50 km/sec, 100 km/sec, 200 km/sec, and 400 km/sec. 
The range of $v_{\rm avg}$ is broadly taken because there are no observations of the velocity of isolated BHs, except for OGLE-2011-BLG-0462/MOA-2011-BLG-101 \citep{sah22, lam22, lam23} that has a transverse velocity of $37.6 \pm 5.1$ km/sec \citep{lam23}.

Also, instead of a fixed speed used in the original model, we consider a kick velocity that follows the Maxwell distribution 
\begin{align}
f(v_{\rm kick}) &= \frac{4 \pi v_{\rm kick}^2}{(2\pi \sigma_{\rm kick}^2)^{3/2}} \exp{\left (-\frac{v_{\rm kick}^2}{2 \sigma_{\rm kick}^2} \right )}, \label{eq:vkick}
\end{align}
where $\sigma_{\rm kick}^2 = \pi \, v_{\rm avg}^2 /8$ is the 1D velocity dispersion which is the same in $x$, $y$, and $z$ directions (i.e., $\sigma_{\rm kick}^2 = \sigma_{x, {\rm kick}}^2 = \sigma_{y, {\rm kick}}^2 = \sigma_{z, {\rm kick}}^2$), and $v_{\rm kick}$ is the magnitude of the 3D kick velocity.
Note that \citet{ros22} showed that the $t_{\rm E}$ distribution of microlensing events due to  BH lenses is insensitive to whether $v_{\rm kick}$ is given by a fixed value or by the Maxwell distribution with the same average value as the fixed value.

\subsubsection{Scale height of BH disk}

Although not considered in \citetalias{kos21} or \citet{lam20}, the BH population should be in different states of dynamical equilibrium than the other stars because of the additional kick velocity.
\citet{tsu18} numerically evaluated this effect by calculating BHs' orbits in a potential that mimics our Galaxy as a function of average BH velocity, and we combine their results with the \citetalias{kos21}'s model\footnote{Ideally, we should repeat such a calculation of orbits under a potential corresponding to the stellar density profile of the \citetalias{kos21} model. However, it is difficult because the \citetalias{kos21} model is not a self-consistent model in terms of dynamics. Keeping consistency with both dynamics and the data fitted in \citetalias{kos21} requires a lot of effort and is beyond the scope of this paper. Nevertheless, combining the results of \citet{tsu18} with the \citetalias{kos21} model should be approximately valid because both models are supposed to reproduce our Galaxy.}.

\begin{figure}
\begin{center}
\includegraphics[width=11cm]{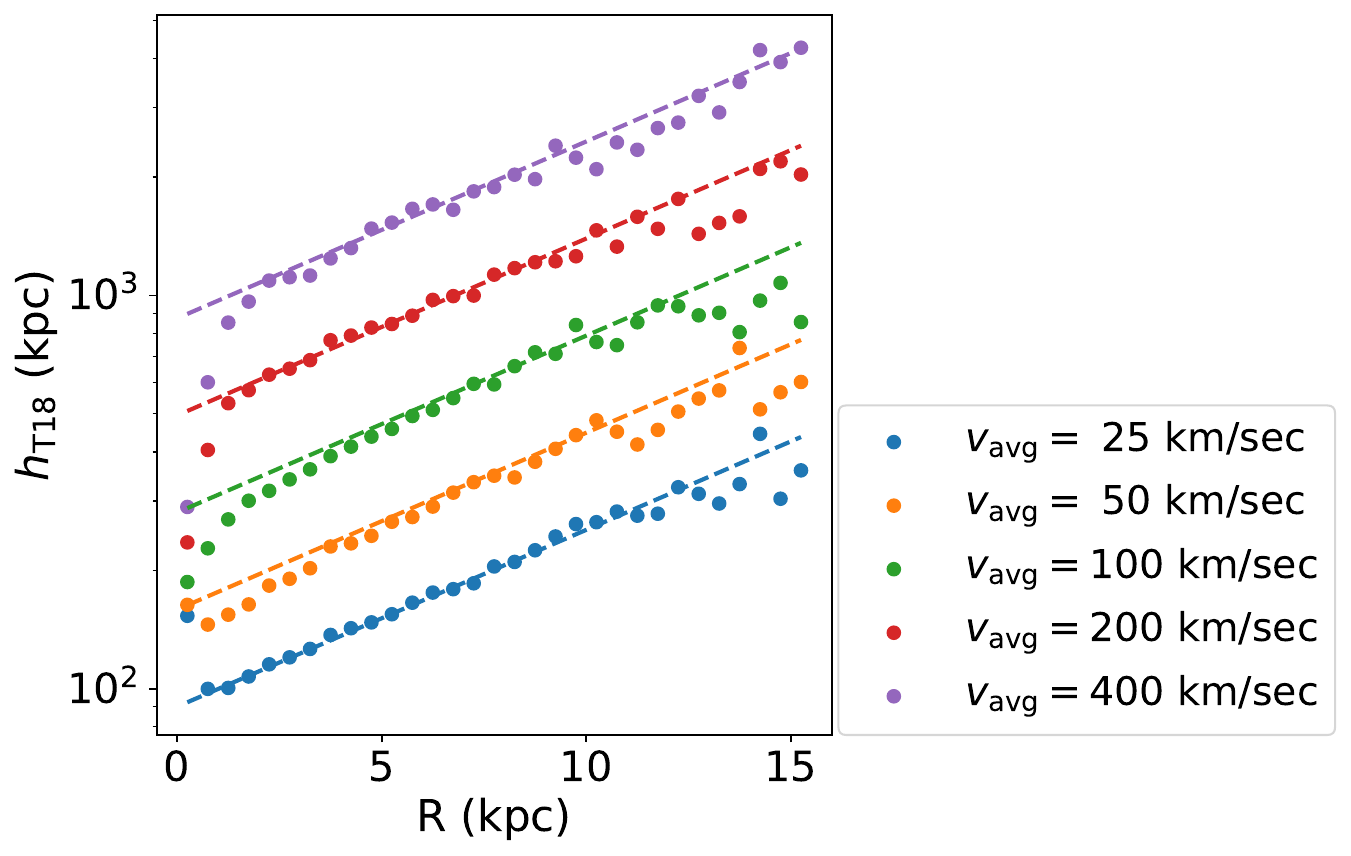}
\caption{
  \label{fig:R_hT18}
Scale height distribution of the BH density profile in the \citet{tsu18}'s numerical model as a function of the radius from the Galactic Center and the average velocity of the BH population.
Dashed lines show our analytical model approximation fitted to the numerical model at $1.25 < R/{\rm kpc} < 8.75$.
}
\end{center}
\end{figure}

We at first see how the scale height in the \citet{tsu18} model depends on the radius from the Galactic Center, $R$, and on the average velocity of the Maxwell distribution of the BH natal kick, $v_{\rm avg}$. Here, the scale height in the \citet{tsu18} model is defined by $h_{\rm T18} \equiv \Sigma_{\rm T18}/(2 \rho_{\rm T18})$, where $\Sigma_{\rm T18}$ is the surface density of the BH distribution in dynamical equilibrium, and $\rho_{\rm T18}$ is the volume density at the disk plane.
Fig. \ref{fig:R_hT18} shows the distribution of $h_{\rm T18}$ as a function of $R$ and $v_{\rm avg}$. We find that $h_{\rm T18}$ at $1.25 < R/{\rm kpc} < 8.75$ is well explained by an analytic formula,
\begin{align}
h_{\rm T18} = 210 \, {\rm pc} \, \exp\left[\frac{R - R_{\odot}}{R_{\rm BH, 0}}\right] \, \left( \frac{v_{\rm avg}} {25 \, {\rm km \, sec^{-1}} }\right)^{\beta_{\rm BH}}, \label{eq:hT18}
\end{align}
where $R_{\rm BH, 0} = 9.66~$kpc, $\beta_{\rm BH} = 0.82$, and $R_{\odot}$ is the distance to the Galactic Center from the Sun and we use $R_{\odot} =$ 8.16 kpc.
The inner region of $R < 1~{\rm kpc}$ was not fitted by this formula, as shown in Fig. \ref{fig:R_hT18}. 
This is because the bulge stars, which have different distributions from disk stars, dominate the area. Here, we do not consider any modifications to the bulge density model and use the original model of \citetalias{kos21} for simplicity.
In Section \ref{sec:mod_bar}, we apply a simple modification to the bulge density model to show that it does not change our discussion about constraints on the kick velocity.
The following modifications are only for the disk density model.

The $v_{\rm avg}$ dependence of $h_{\rm T18}$ can be interpreted as due to the vertical velocity dispersion dependence of the scale height, which is well known for the stellar component \citep[e.g.,][]{van11}.
We incorporate this effect in the disk model by \citetalias{kos21} as
\begin{align}
h_{{\rm BH}, i} = h_{0, i} \, g(R) \,\left( \frac{\sigma_{z, i, {\rm BH}} (R)} {\sigma_{z, i} (R)}\right)^{\beta_{\rm BH}}, \label{eq:h_BH}
\end{align}
with the vertical velocity dispersion of BHs enhanced due to the kick, 
\begin{align}
\sigma_{z, i, {\rm BH}} (R) = \sqrt{\sigma_{z, i} (R)^2 + \pi \, v_{\rm avg}^2 /8},
\end{align}
where $i$ takes 1 to 8 to denote each of the seven thin disks with different ages  ($i = $1, 2, ... 7) and one thick disk ($i = 8$) in the \citetalias{kos21} model, $h_{0, i}$ is the scale height of $i$th disk that is given in Table 1 of \citetalias{kos21} as $z_{\rm d, \odot}$, and $\sigma_{z, i} (R)$ is the vertical velocity dispersion of stars in $i$th disk which is given by 
\begin{equation}
 \sigma_{z, i} (R) = 
  \left\{
   \begin{array}{ll}
     {\rm 24.4 \, km\, sec^{-1}} \, \left ( \frac{T_i + T_{\rm min}}{T_{\rm max} + T_{\rm min}} \right)^{0.77} \exp \left[ - \frac{R - R_{\odot}}{\rm 5.9 \, kpc} \right]  & (i = 1, 2, ... 7)\\
     {\rm 49.2 \, km\, sec^{-1}} \exp \left[ - \frac{R - R_{\odot}}{\rm 9.4 \, kpc} \right] & (i = 8), \\
    \end{array}
    \right.
  \label{eq:sigz}
\end{equation}
where $T_{\rm min} = 0.01$ Gyr, $T_{\rm max} = 10$ Gyr, and $T_i$ is the mean age of the stars in $i$th thin disk given by 0.08 Gyr, 0.59 Gyr, 1.52 Gyr, 2.52 Gyr, 4.07 Gyr, 6.07 Gyr and 8.66 Gyr for $i = 1$ to 7, respectively, which reflects the assumed star formation rate for the thin disk of $\propto \exp \left[ T/(7~{\rm Gyr}) \right]$ \citep{bov17}.

The remaining factor in Eq. (\ref{eq:h_BH}) not yet explained is $g (R)$.
This is introduced to smoothly connect the $h_{\rm T18}$ formula depending on $R$ by $\exp [(R - R_{\odot})/R_{\rm BH, 0}]$ to the scale height of \citetalias{kos21}'s disk model that is constant for $R$.
The scale height in \citet{tsu18}, $h_{\rm T18}$, increases with $R$ because they assumed that BHs have a uniform velocity dispersion regardless of their positions, which results in a larger scale height at the outer part of the Galaxy where the gravity is weaker.  On the other hand, \citetalias{kos21} introduced a model for the velocity dispersion of stars that decreases with $R$, which is supported by the Gaia data \citep{kat18}.  Therefore, we need a formula to give the scale height of BHs that bridges two regimes which have different $R$-dependence of the scale height: the regime where the stellar velocity dispersion $\sigma_{z,i}(R)$ is dominant and the regime where the kick velocity dispersion $\pi \, v_{\rm avg}^2 /8$ is dominant.
We use 
\begin{align}
g (R) = \exp\left[\frac{R - R_{\odot}}{R_{\rm BH} (R)}\right],  \label{eq:gR}
\end{align}
where
\begin{align}
R_{\rm BH} (R) = R_{\rm BH, 0}  \left(1 + \frac{\sigma_{z, i}(R)^2}{\pi \, v_{\rm avg}^2 /8} \right), \label{eq:RBH}
\end{align}
to connect smoothly the regime of large kick velocity (i.e., $\pi v_{\rm avg}^2 \gg \sigma_{z,i}$) and that of small kick velocity (i.e., $\pi v_{\rm avg}^2 \ll \sigma_{z,i}$).
Although this is a crude attempt to include a dependence on $R$, we show in Section \ref{sec:unc_gR} that our main results of the constraint on kick velocity are not affected by how this is connected.


\begin{figure}
\begin{center}
\includegraphics[width=9cm]{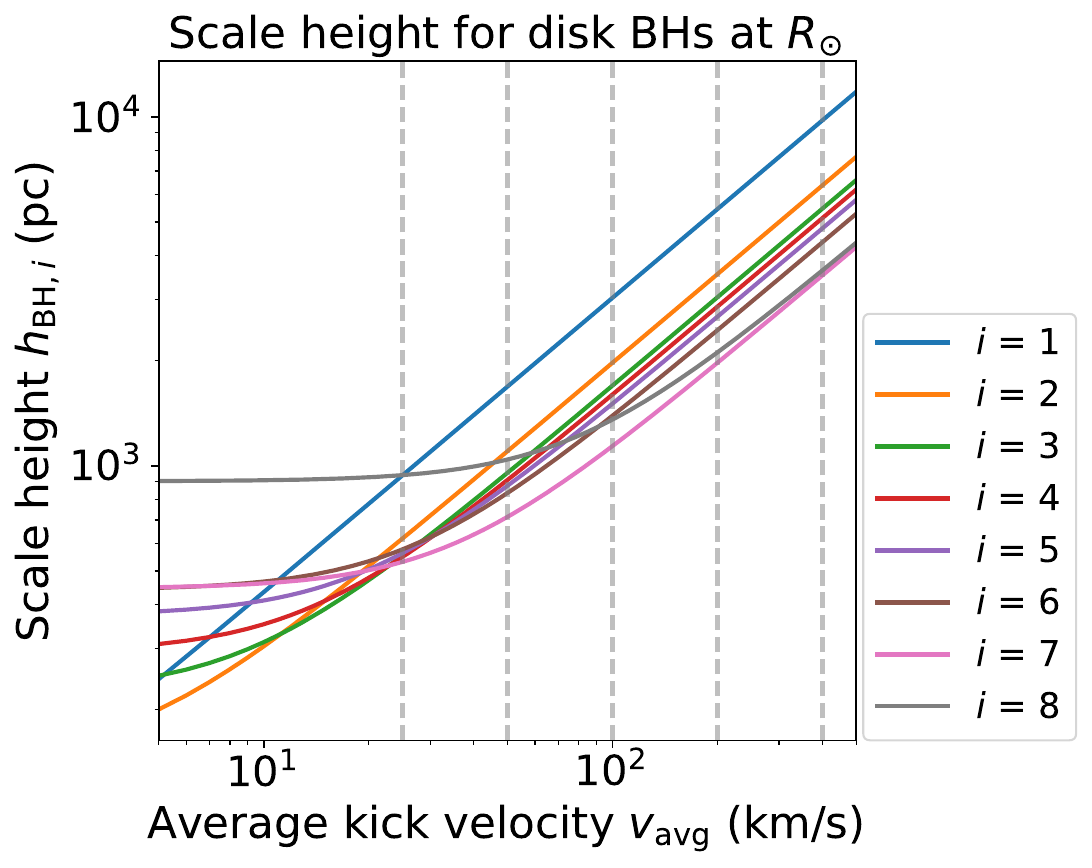}
\caption{
  \label{fig:vavg_hBH}
Our scale height model for disk black holes as a function of the average velocity of the Maxwell distribution of the BH natal kick.
Different color indicates different disk components ($i= 1$--7 for thin disk and $i = 8$ for thick disk). Five vertical dashed gray lines indicate the considered $v_{\rm avg}$ values (25 km/sec, 50 km/sec, 100 km/sec, 200 km/sec, and 400 km/sec) in this paper.
}
\end{center}
\end{figure}

Fig. \ref{fig:vavg_hBH} shows the scale height for disk BHs (i.e., Eq. \ref{eq:h_BH}) as a function of the average kick velocity $v_{\rm avg}$ for each of $i$th disk component at $R = R_{\rm \odot}$. 
At the high end of $v_{\rm avg}$, all the components have lost their original velocity dispersion information and have the same velocity dispersion due to the kick velocity.
In such a region, a smaller $i$ (younger) component has a larger scale height because its surface density is less massive than older component disks, and therefore its gravity is smaller.

\subsubsection{Surface density of BH disk}
\begin{figure}
\begin{center}
\includegraphics[width=8cm]{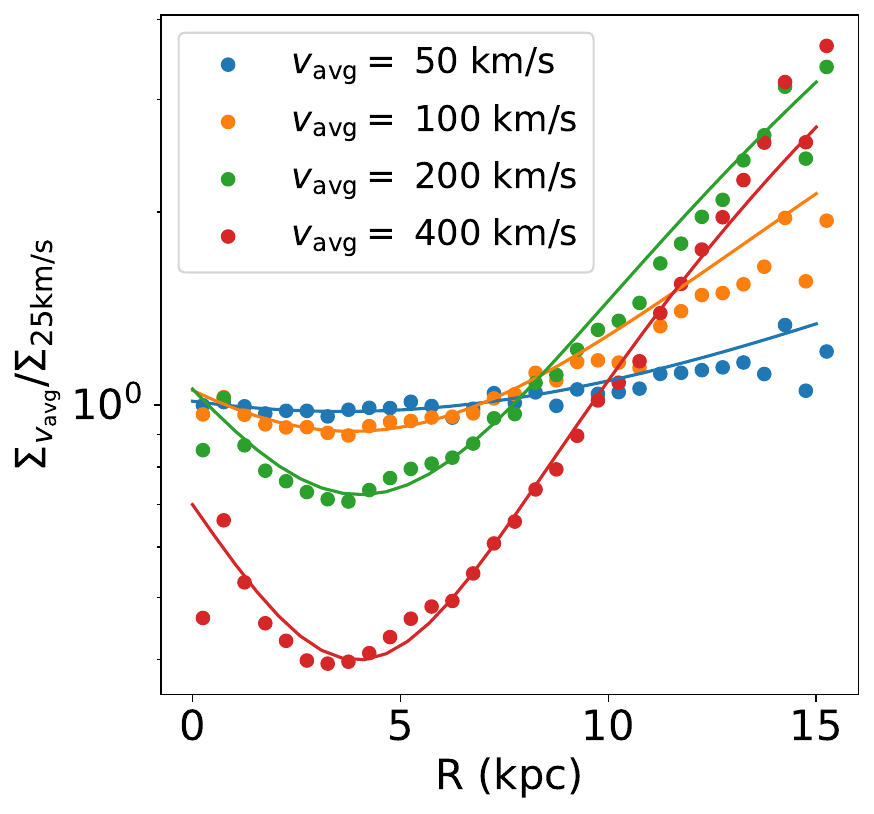}
\caption{
  \label{fig:Sigma_vavg}
Dots show the surface density with $v_{\rm avg} =$ 50 km/sec, 100 km/sec, 200 km/sec, and 400 km/sec as a function of $R$ relative to the one with $v_{\rm avg} = 25$ km/sec from the numerical model by \citet{tsu18}. The solid lines show the second-order polynomial curves fitted to each of the three in $0.75 < R/{\rm kpc} < 8.25$.
}
\end{center}
\end{figure}

\begin{deluxetable}{rlccccccccccccccccc}
\tablecaption{Comparison of properties between stars and BHs. \label{tab:mod2BH}}
\tablehead{
\colhead{Property}  & \colhead{Location}  & Stars & BHs & Refer to
}
\startdata
Velocity        & Disk \& Bulge & ${\bm v}$ & ${\bm v} + {\bm v_{\rm kick}}$ & Eq.(\ref{eq:vkick})\\
Scale height    & Disk          & $h_{0, i}$ & $h_{0, i} \, g(R) \,\left( \frac{\sigma_{z, i, {\rm BH}} (R)} {\sigma_{z, i} (R)}\right)^{\beta_{\rm BH}}$ & Eq.(\ref{eq:h_BH})  \\
Surface density & Disk          & $\Sigma_0 (R)$  & $\Sigma_0 (R) \, \eta(R ; v_{\rm avg})$ & Eq.(\ref{eq:SigBH})\\
\enddata
\end{deluxetable}

In addition to the scale height dependence on the kick velocity, the numerical model by \citet{tsu18} also shows the dependence of the surface density on the kick velocity.
Dots in Fig. \ref{fig:Sigma_vavg} show the surface density in the numerical models by \citet{tsu18} for $v_{\rm avg} =$ 50 km/sec, 100 km/sec, 200 km/sec, and 400 km/sec relative to the one with $v_{\rm avg} = 25$ km/sec.
We found that a second-order polynomial curve (shown in Fig. \ref{fig:Sigma_vavg}) can well-fit each of the relative surface densities.
Thus, we use
\begin{align}
\Sigma_{\rm BH} (R ; v_{\rm avg}) = \Sigma_0 (R) \, \eta(R ; v_{\rm avg}) \label{eq:SigBH}
\end{align}
with $\eta(R ; v_{\rm avg}) = a + b \, (R/{\rm kpc}) + c \, (R/{\rm kpc})^2$ for the surface density of black hole population with the average kick velocity of $v_{\rm avg}$.
Here, $\Sigma_0 (R)$ is the original surface density of the \citetalias{kos21} model and we determined the coefficients $a$, $b$, and $c$ in $\eta(R ; v_{\rm avg})$ based on Fig. \ref{fig:Sigma_vavg}; $(a, b, c) = (1, 0, 0)$, $(1.013, -0.020, 0.003)$, $(1.054, -0.075, 0.010)$, $(1.059, -0.166, 0.021)$ and $(0.699, -0.151, 0.019)$ for $v_{\rm avg} =$ 25 km/sec, 50 km/sec, 100 km/sec, 200 km/sec, and 400 km/sec, respectively.


All modifications applied to the BH population described in this section are summarized in Table \ref{tab:mod2BH}.


\section{Influence on microlensing distributions}
As described in Section \ref{sec:intro}, the two commonly measurable microlens parameters in the case of BH events are the Einstein radius crossing time $t_{\rm E}$ and the microlens parallax $\pi_{\rm E}$.
Because $t_{\rm E} \propto \sqrt{M_{\rm L}}$, BH events tend to have large values of $t_{\rm E}$.
The microlens parallax $\pi_{\rm E}$ can be measured in such a long timescale event through the microlens parallax effect due to the Earth's orbital motion, and it is defined as $\pi_{\rm E} \equiv \pi_{\rm rel}/\theta_{\rm E}$.

Here, we consider the influence of changing kick velocity on the $t_{\rm E}$ distribution and the two-dimensional distribution of $t_{\rm E}$ and $\pi_{\rm E}$.
In this work, we use a modified version of \texttt{genulens} \citep{kosran22} to simulate microlensing events following the model described in Section \ref{sec:model}.

\subsection{Event rate as a function of $t_{\rm E}$}

\subsubsection{Influence on disk BH event rate}
\begin{figure}
\begin{center}
\includegraphics[width=16cm]{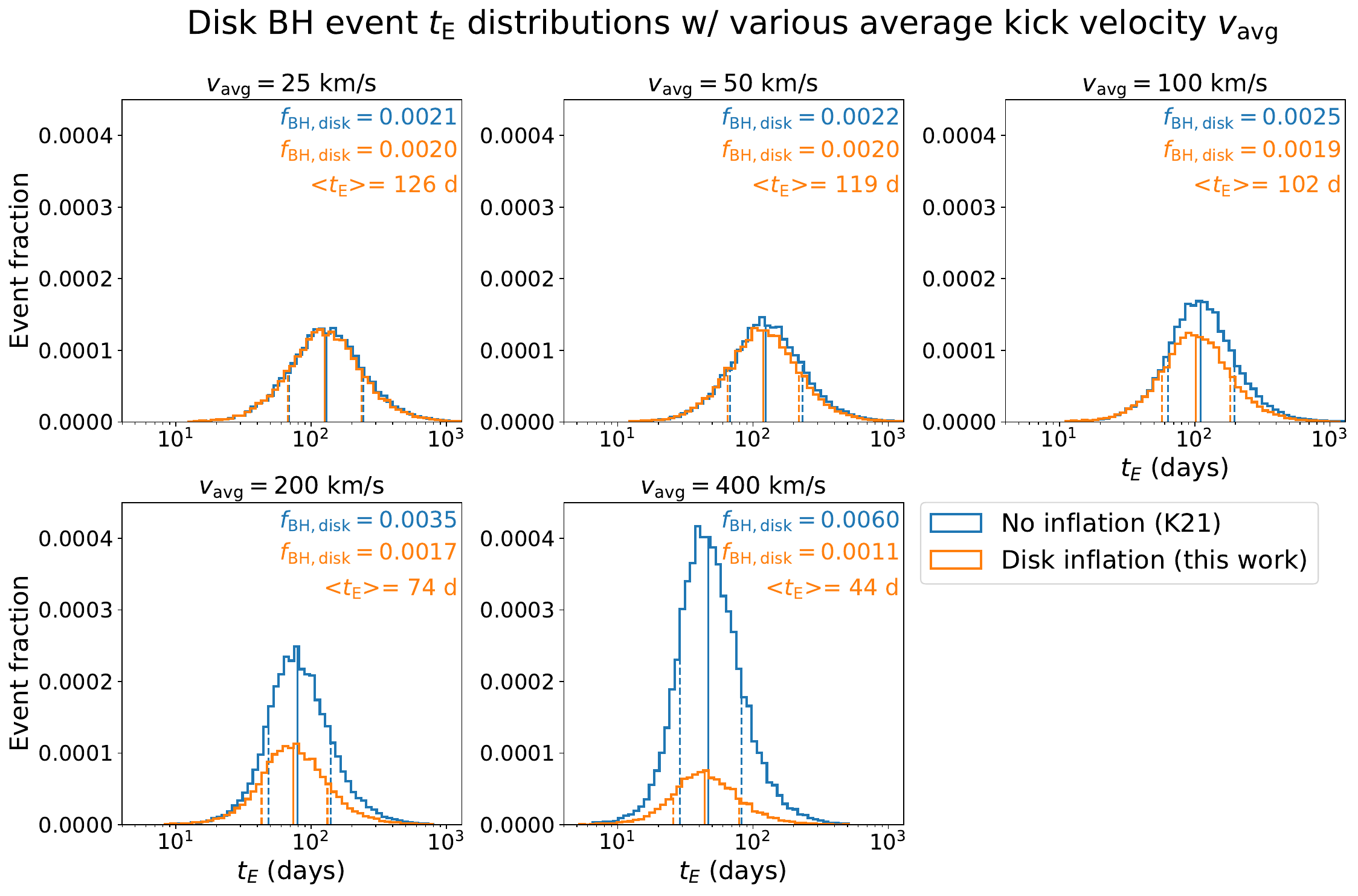}
\caption{
  \label{fig:diskBH_tEs}
Event fraction of disk BH events as a function of $t_{\rm E}$ for different average kick velocity $v_{\rm avg}$.
The event fraction is relative to all the simulated $3 \times 10^7$ microlensing events toward  $(l, b) = (-0.14, -1.62)$, the event coordinate of the isolated stellar-mass BH event OGLE-2011-BLG-0462/MOA-2011-BLG-101.}
\end{center}
\end{figure}

Fig. \ref{fig:diskBH_tEs} shows the $t_{\rm E}$ distribution of black hole lens events as a function of the average kick velocity $v_{\rm avg}$.
We use a Monte Carlo simulation of $3 \times 10^7$ realizations of microlensing events toward $(l, b) = (-0.14, -1.62)$, i.e., the event coordinate of the isolated stellar-mass BH event OGLE-2011-BLG-0462/MOA-2011-BLG-101 \citep[hereafter OB110462,][]{sah22, lam22}, and the Fig. \ref{fig:diskBH_tEs} shows the fraction of disk BH events out of all simulated events.
In each $v_{\rm avg}$ panel, the blue histogram shows the distribution when the kick velocity is added to the BH velocities but the scale height or the surface density does not change (same as the treatment by previous studies such as \citealt{lam20} or \citetalias{kos21}). The orange histogram shows the distribution when the scale height and the surface density are also modified according to the added kick velocity.

Because $t_{\rm E} \propto \mu_{\rm rel}^{-1}$, $t_{\rm E}$ gets shorter as $v_{\rm avg}$ gets faster.
This generally results in a higher event rate because the microlensing event rate is proportional to $\mu_{\rm rel} \theta_{\rm E}$, the area of the sky swept by the Einstein ring of an event per unit time.
This is why the blue histogram shows a higher event fraction with faster $v_{\rm avg}$.
On the other hand, the orange histogram shows the opposite trend in terms of the event rate compared to the blue histogram although the shape of the $t_{\rm E}$ distribution itself is similar.
The difference between the blue and orange histograms in the event fraction is larger when $v_{\rm avg}$ is faster.
The difference is relatively small when $v_{\rm avg} \leq 100~{\rm km/sec}$ because the additional velocity due to the kick is yet comparable to the progenitors' original velocity, whereas it gets more prominent at $v_{\rm avg} \geq 200~{\rm km/sec}$.
It predicts two times fewer BH events at $v_{\rm avg} = 200$ km/sec and $\sim 6$ times fewer BH events at $v_{\rm avg} = 400$ km/sec when we consider the inflation of the BH disk due to the kick velocity.
This is because the BH disk becomes thicker with a faster kick velocity as shown in Fig. \ref{fig:vavg_hBH}, and its abundance at a low latitude field (e.g., toward the Galactic bulge) becomes relatively less dominant.
Therefore, we would end up overestimating the BH event fraction especially when the kick velocity is fast, unless we properly consider the scale height inflation due to the kick.

\subsubsection{Comparison with the OGLE-IV $t_{\rm E}$ distribution} \label{sec:comp2ogle}
\begin{figure}
\begin{center}
\includegraphics[width=16cm]{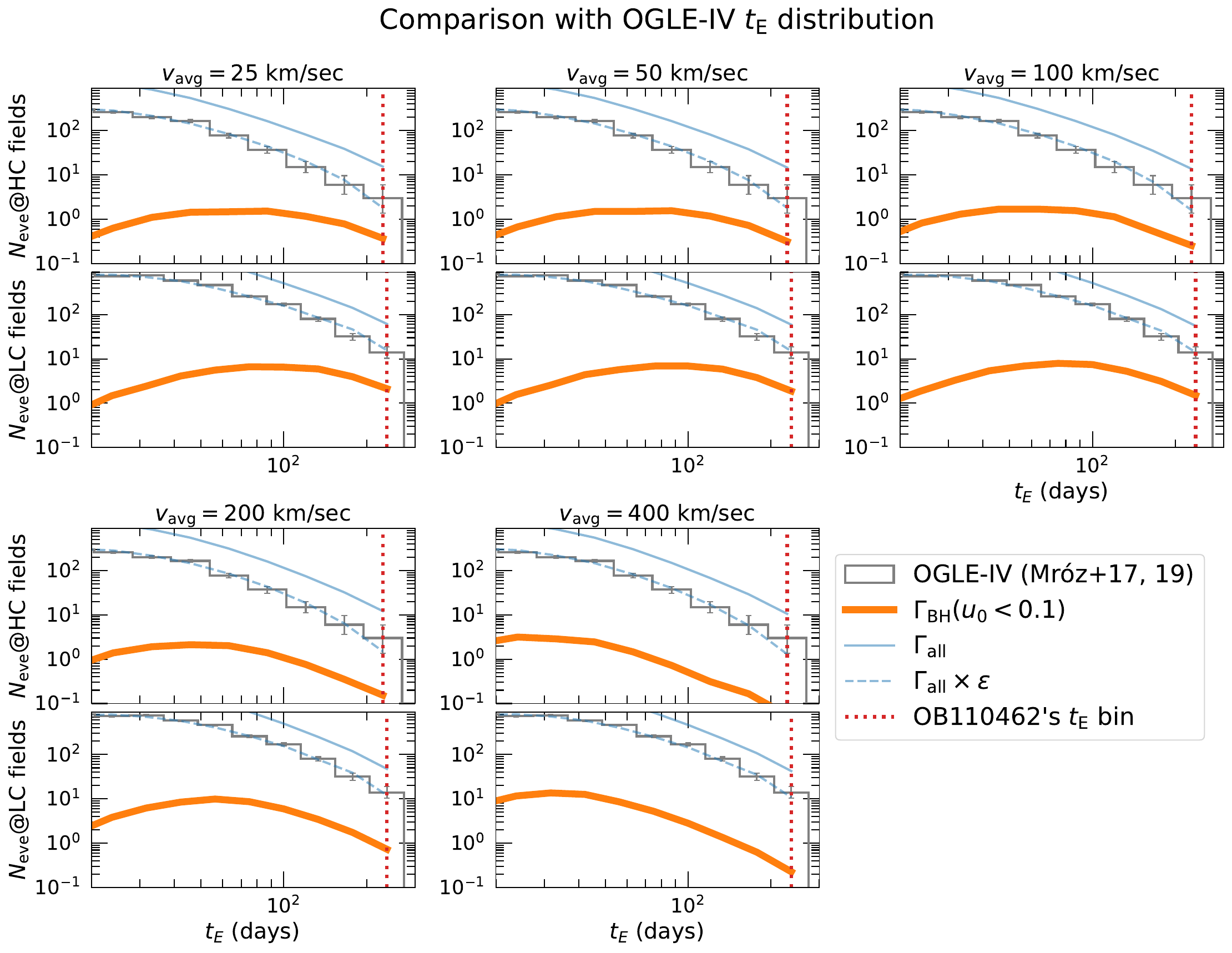}
\caption{
  \label{fig:tE_OGLEs}
Observational $t_{\rm E}$ distributions in the OGLE-IV high-cadence \citep[top of each panel,][]{mro17} and low-cadence \citep[bottom of each panel,][]{mro19} fields compared with model event rate calculated for different average kick velocity $v_{\rm avg}$. The dashed blue line shows the event rate multiplied by the OGLE-IV detection efficiency which fits the observational distributions well. The solid blue line shows the event rate before multiplying the detection efficiency, corresponding to the number of events occurring during the OGLE-IV survey. The solid orange line shows that of black hole lenses but for $u_0 < 0.1$.
}
\end{center}
\end{figure}

\citet{mro17} analyzed microlensing events discovered in the OGLE-IV high cadence fields between 2010 and 2015, while \citet{mro19} analyzed microlensing events in the OGLE-IV low cadence fields between 2010 and 2017. They provided $t_{\rm E}$ distributions and detection efficiencies (private communication with P. Mr{\'o}z) in both fields.

Fig. \ref{fig:tE_OGLEs} shows the comparison of the OGLE-IV $t_{\rm E}$ distributions with the model distributions calculated with different average kick velocity $v_{\rm avg}$.
With each $v_{\rm avg}$ value, we performed a Monte Carlo simulation to have a model $t_{\rm E}$ distribution in the OGLE-IV survey fields. We then multiplied the detection efficiency by the model $t_{\rm E}$ distribution, and we normalized the distribution by the total number of OGLE-IV events detected to have the blue dashed curves, $\Gamma_{\rm all} (t_{\rm E}) \epsilon (t_{\rm E})$, in Fig. \ref{fig:tE_OGLEs} (see section 4.1.5 of \citetalias{kos21} for more details of this calculation). In this way, the original model $t_{\rm E}$ distribution, shown in blue solid curves, is also normalized.

Fig. \ref{fig:tE_OGLEs} shows that the observed distribution (gray histograms) can be well reproduced by the model event rate multiplied by the detection efficiency (blue dashed curve) regardless of the $v_{\rm avg}$ value. This indicates all the considered $v_{\rm avg}$ values are acceptable in terms of consistency with the OGLE $t_{\rm E}$ distribution.
However, we know that there is one BH event, OB110462 \citep{sah22, lam22}, that occurred during the OGLE-IV survey period analyzed by the Mr{\'o}z et al. papers shown as red dotted lines. This fact enables us to constrain the kick velocity, as discussed in detail in Section \ref{sec:disc_vave}.

\subsection{Distribution of $t_{\rm E}$ vs $\pi_{\rm E}$} \label{sec:tEpiE}
\begin{figure}
\begin{center}
\includegraphics[width=16cm]{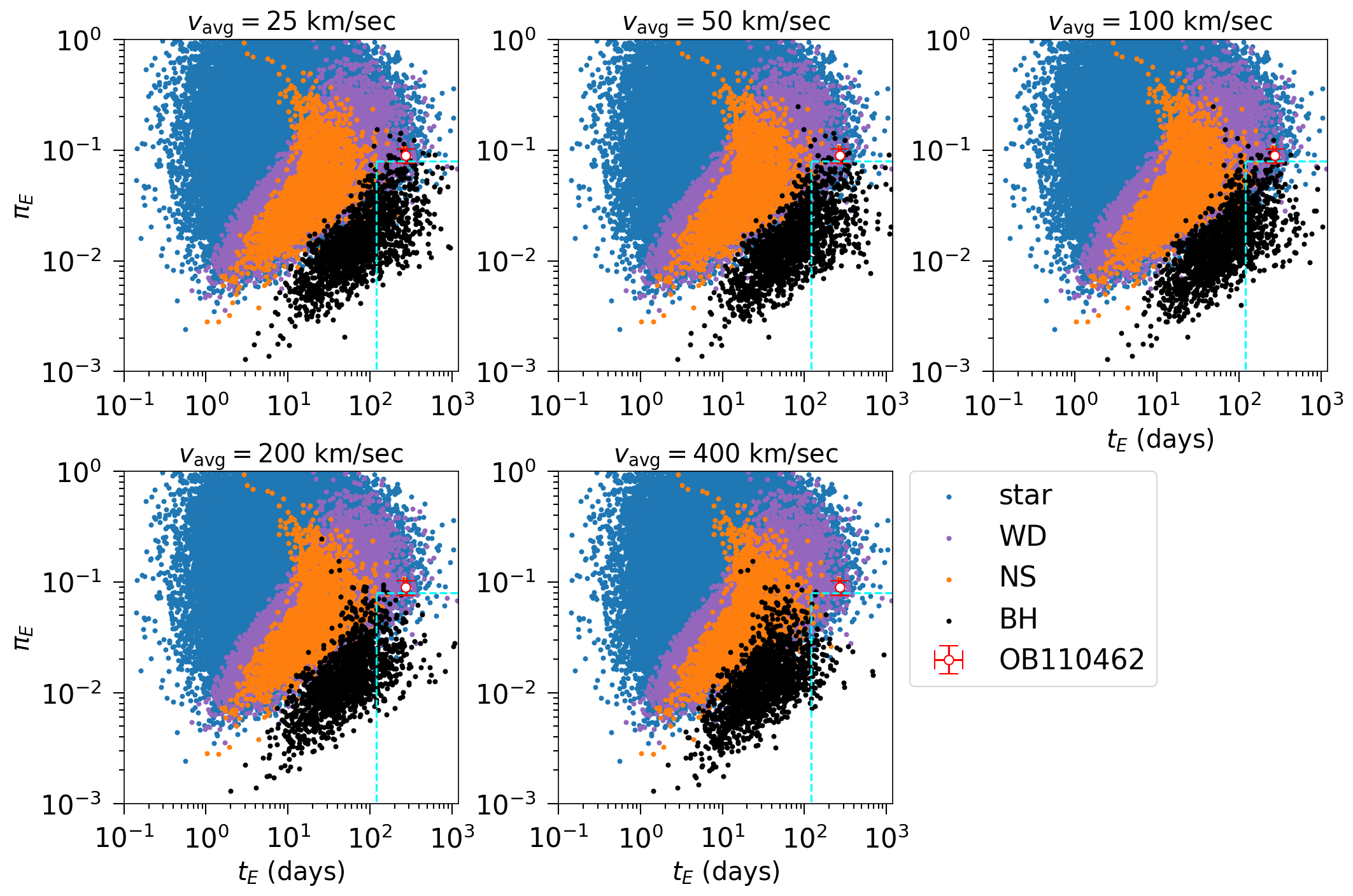}
\caption{
  \label{fig:tE_piEs}
$t_{\rm E}$ vs $\pi_{\rm E}$ distribution of $2 \times 10^5$ events randomly generated following our Galactic model with $v_{\rm avg} =$ 25 km/sec, 50 km/sec, 100 km/sec, 200 km/sec, and 400 km/sec.
The cyan dashed lines indicate the region of $t_{\rm E} > 120$ days and $\pi_{\rm E} < 0.08$ suggested by \citet{lam20} as a selection criterion for BH events. Note that \citet{lam20} applied a single BH kick velocity of 100 km/sec.
}
\end{center}
\end{figure}

A similar approach can be done on the $t_{\rm E}$ vs $\pi_{\rm E}$ plane.
\citet{lam20} showed that $\sim 85\%$ of the lenses are due to BHs in events with $t_{\rm E} > 120$ days and $\pi_{\rm E} < 0.08$ when the kick velocity is 100 km/sec for BHs.
This is also confirmed with our model as shown in the $v_{\rm avg} = 100$ km/sec panel in Fig. \ref{fig:tE_piEs}, where the cyan dashed lines indicate the region of $t_{\rm E} > 120$ days and $\pi_{\rm E} < 0.08$. The figure also shows the same distribution with different $v_{\rm avg}$ values. This indicates that the $t_{\rm E}$ distributions of the BH population are similar in $v_{\rm avg} =$ 25 km/sec, 50 km/sec, and 100 km/sec, but the distribution shifts to the shorter $t_{\rm E}$ side as $v_{\rm avg}$ exceeds 100 km/sec. This is because the additional kick velocity becomes more dominant than the original celestial velocity at $v_{\rm avg} >$ 100 km/sec.

The BH population (black dots) is not distributed around the $(t_{\rm E}, \pi_{\rm E})$ value observed for OB110462 when $v_{\rm avg} =$ 200 km/sec and 400 km/sec, which implies these high kick velocity values are unlikely. The rare position of OB110462 among the black dots is due to its close distance from the Sun of $D_{\rm L} = 1.72^{+0.32}_{-0.23}~{\rm kpc}$ \citep{lam23}, significantly closer than the typical ($\sim 2\,\sigma$) range of $D_{\rm L}$ for BH lenses in our model of $\sim 3$~kpc to $\sim 9$~kpc. We further discuss this event in Section \ref{sec:disc_vave}.

\section{Implication on average kick velocity from OGLE-2011-BLG-0462} \label{sec:disc_vave}

{\tabcolsep = 1.2mm
\begin{deluxetable}{lccccccccccccccccccccccccccccccc}
\tabletypesize{\footnotesize}
\tablecaption{Estimated number of black hole events like OGLE-2011-BLG-0462 occurred during the OGLE-IV survey by \citet{mro17} and \citet{mro19}. \label{tab-num462}}
\tablehead{            
Model\tablenotemark{a} & \multicolumn{14}{c}{Average kick velocity $v_{\rm avg}$}\\
                       & \multicolumn{2}{c}{25 km/sec} & & \multicolumn{2}{c}{50 km/sec} & & \multicolumn{2}{c}{100 km/sec} & & \multicolumn{2}{c}{200 km/sec} & & \multicolumn{2}{c}{400 km/sec}\\
                      \cline{2-3} \cline{5-6} \cline{8-9} \cline{11-12} \cline{14-15}
                      & \multicolumn{2}{c}{\bm {$\bm{N_{\rm exp, 0462}}$}} & & \multicolumn{2}{c}{{$\bm{N_{\rm exp, 0462}}$}} & & \multicolumn{2}{c}{{$\bm{N_{\rm exp, 0462}}$}} & & \multicolumn{2}{c}{{$\bm{N_{\rm exp, 0462}}$}} & & \multicolumn{2}{c}{{$\bm{N_{\rm exp, 0462}}$}}\\
                      & ($\Gamma_{\rm BH}$ & $P_{\rm BH}$)    & & ($\Gamma_{\rm BH}$ & $P_{\rm BH}$) & & ($\Gamma_{\rm BH}$ & $P_{\rm BH}$) & & ($\Gamma_{\rm BH}$ & $P_{\rm BH}$) & & ($\Gamma_{\rm BH}$ & $P_{\rm BH}$)
}
\startdata
Fiducial                  & \multicolumn{2}{c}{\bf 0.26} & & \multicolumn{2}{c}{\bf 0.19}    & & \multicolumn{2}{c}{\bf 0.095}    & & \multicolumn{2}{c}{\bf 0.020}    & & \multicolumn{2}{c}{\bf 0.0018} \\
                          &       (2.5 & 0.10)           & &       (2.2 & 0.086)             & &       (1.8 & 0.054)              & &      (0.88 & 0.023)              & &     (0.28 & 0.0065)            \\\hline
Inflate-bar               & \multicolumn{2}{c}{\bf 0.26} & & \multicolumn{2}{c}{\bf 0.19}    & & \multicolumn{2}{c}{\bf 0.094}     & & \multicolumn{2}{c}{\bf 0.020}   & & \multicolumn{2}{c}{\bf 0.0023} \\
                          &       (2.5 & 0.10)           & &       (2.2 & 0.087)             & &       (1.7 & 0.056)              & &      (0.74 & 0.027)              & &     (0.15 & 0.015)             \\\hline
{\"O}zel-MF               & \multicolumn{2}{c}{\bf 0.29} & & \multicolumn{2}{c}{\bf 0.21}    & & \multicolumn{2}{c}{\bf 0.097}    & & \multicolumn{2}{c}{\bf 0.020}    & & \multicolumn{2}{c}{\bf 0.0019} \\
                          &       (1.9 & 0.16)           & &       (1.6 & 0.13)              & &       (1.2 & 0.081)              & &      (0.58 & 0.034)              & &     (0.19 & 0.010)             \\\hline
$P_{\rm evo, BH} = 1$     & \multicolumn{2}{c}{\bf 0.44} & & \multicolumn{2}{c}{\bf 0.32}    & & \multicolumn{2}{c}{\bf 0.15}     & & \multicolumn{2}{c}{\bf 0.031}    & & \multicolumn{2}{c}{\bf 0.0026} \\
                          &       (2.8 & 0.16)           & &       (2.5 & 0.13)              & &       (1.8 & 0.081)              & &      (0.89 & 0.034)              & &     (0.26 & 0.010)             \\\hline
$R_{\rm BH}=R_{\rm BH,0}$ & \multicolumn{2}{c}{\bf 0.28} & & \multicolumn{2}{c}{\bf 0.19}    & & \multicolumn{2}{c}{\bf 0.096}    & & \multicolumn{2}{c}{\bf 0.020}    & & \multicolumn{2}{c}{\bf 0.0019} \\
                          &       (2.6 & 0.11)           & &       (2.2 & 0.087)             & &       (1.8 & 0.054)              & &      (0.90 & 0.023)              & &     (0.28 & 0.0066)            \\\hline
$R_{\rm BH}=\infty$       & \multicolumn{2}{c}{\bf 0.24} & & \multicolumn{2}{c}{\bf 0.17}    & & \multicolumn{2}{c}{\bf 0.079}    & & \multicolumn{2}{c}{\bf 0.016}    & & \multicolumn{2}{c}{\bf 0.0013} \\
                          &       (2.5 & 0.099)          & &       (2.1 & 0.082)             & &       (1.6 & 0.050)              & &      (0.79 & 0.020)              & &     (0.24 & 0.0056)            \\\hline
\enddata
\tablenotetext{a}{Fiducial model, described in Section \ref{sec:model}, uses the \citet{lam20}'s BH IFMR and considers the disk scale height inflation due to the natal kick.\\
Inflate-bar model is the fiducial model but with a bar scale height inflation due to the natal kick. \\
{\"O}zel-MF model is the fiducial model but uses the \citet{oze10}'s Gaussian BH IFMR. \\
$P_{\rm evo, BH} = 1$ model is the {\"O}zel-MF model but with $P_{\rm evo, BH} = 1$ for the probability of evolving into BHs for stars with $M_{\rm ini} > 15\,M_{\odot}$. \\
$R_{\rm BH}=R_{\rm BH,0}$ model is the fiducial model but uses $R_{\rm BH}=R_{\rm BH,0}$ in Eq. (\ref{eq:RBH}). \\
$R_{\rm BH}=\infty$ model is the fiducial model but uses $R_{\rm BH}=\infty$ in Eq. (\ref{eq:RBH}), i.e., uses $g (R) = 1$.
}
\tablecomments{$\Gamma_{\rm BH}$ refers to $\Gamma_{\rm BH} (u_0 < 0.1, t_{\rm E} > 200~{\rm days})$, $P_{\rm BH}$ refers to $P_{\rm BH} (\pi_{\rm E} > 0.06 \,|\, t_{\rm E, OB110462})$, and $N_{\rm exp, 0462}$ refers to the expected number of both happening.}
\end{deluxetable}
}

\begin{figure}
\begin{center}
\includegraphics[width=8cm]{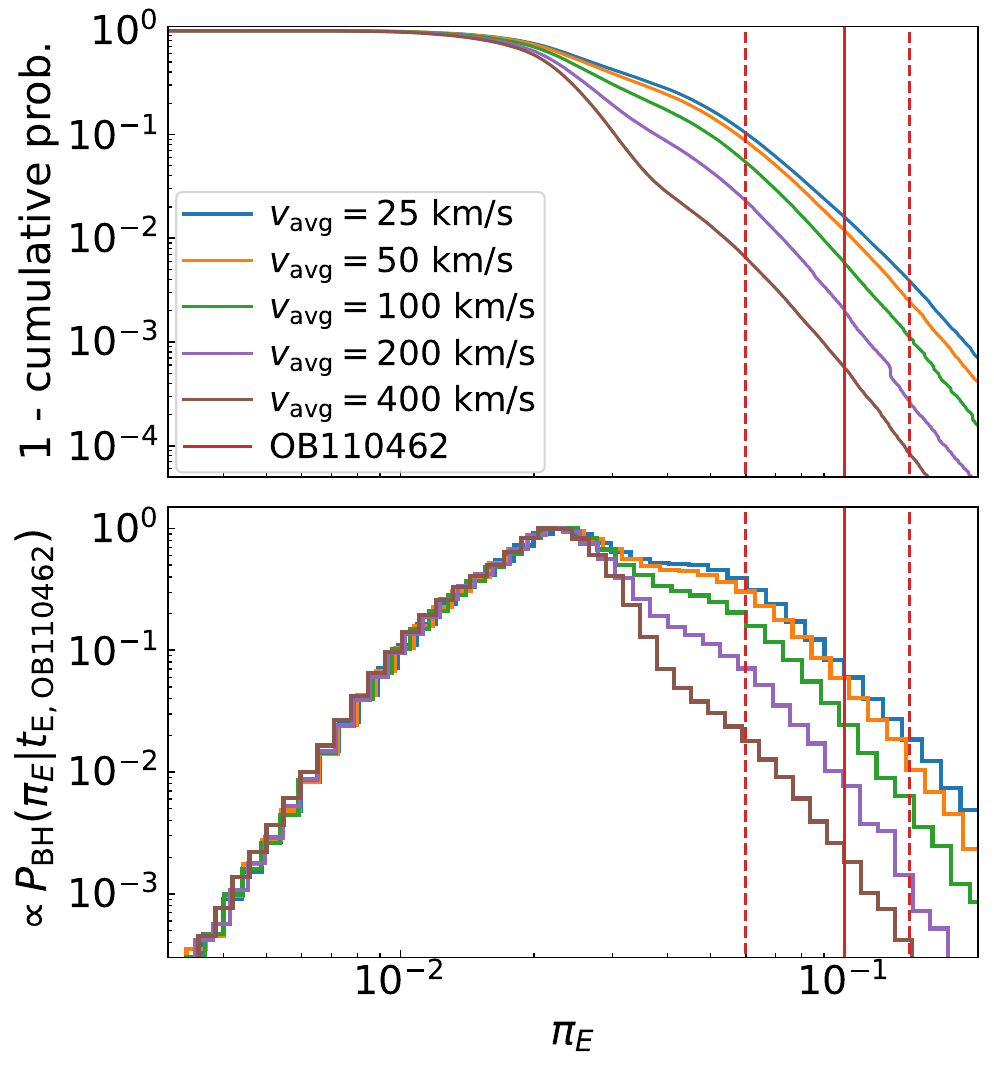}
\caption{
  \label{fig:piE462}
Cumulative probability (top) and relative probability (bottom) distributions of $\pi_{\rm E}$ for BH-lens events with $259 < t_{\rm E}/{\rm days} < 282$ ($1~\sigma$ uncertainty region of $t_{\rm E}$ for OB110462) when $v_{\rm avg}$ is 25 km/sec, 50 km/sec, 100 km/sec, 200 km/sec, and 400 km/sec. 
The relative probabilities in the bottom panel are normalized by the peak probability of each.
The vertical red lines are the median (solid line) and the 3$\sigma$ range (dashed lines) of $\pi_{\rm E}$ of OB110462 estimated by \citet{lam23}.
}
\end{center}
\end{figure}

The only unambiguous isolated BH event to date, OGLE-2011-BLG-0462/MOA-2011-BLG-101 (OB110462), resides in so unique parameter space \citep[$t_{\rm E} = 276 \pm 6$ days, $u_0 = 0.05 \pm 0.01$, $\pi_{\rm E} = 0.10 \pm 0.01$;][]{lam23} that this single event can be used to obtain some restrictions on the kick velocity.
\citet{lam23} also pointed out the rareness of this event but concluded that understanding the selection effect is needed for further discussion.
Indeed, the selection effect is not well understood for this event. Nevertheless, in this section, we attempt to compare the number of BH events expected by a model with the actual number under some plausible assumptions.

The lens object of OB110462 was identified as a BH thanks to the astrometric microlensing measurements using Hubble Space Telescope (HST) data \citep{sah22, lam22, mro22, lam23}. The HST data were observed in a series of HST follow-up programs by Sahu et al. that aimed to discover the first isolated stellar-mass BH via astrometric microlensing.
The HST programs monitored five events selected from those alerted between 2009 and 2011 by the OGLE \footnote{https://ogle.astrouw.edu.pl/ogle4/ews/ews.html} and MOA \footnote{https://www.massey.ac.nz/~iabond/moa/alerts/} alerting systems with $t_{\rm E} > 200$ days and high magnifications \citep{sah09, lam22S}.
The actual size of the total sample that Sahu et al looked at to select the five events is unknown, but it may be assumed to be smaller than the total events alerted from 2009 to 2011.

On the other hand, the OGLE-IV sample plotted in Fig. \ref{fig:tE_OGLEs} was systematically selected from the OGLE-IV high cadence fields between 2010 and 2015 by \citet{mro17} and from the OGLE-IV low cadence fields between 2010 and 2017 by \citet{mro19}.
The combined sample covers the entire OGLE-IV bulge fields for 6 years, and the sample size should be larger than the one that Sahu et al. looked at.
Thus, we here consider how many BH events are expected in the OGLE-IV sample by Mr{\'o}z et al. with $u_0 < 0.1$ and $t_{\rm E} > 200~{\rm days}$ and compare the number with the actual number, at least one. 
The range $u_0 < 0.1$ is determined to meet the requirement of high magnifications in the event selection by the HST programs.
The interpretation of ``high magnifications" is somewhat arbitrary, and we have conservatively taken $u_0 < 0.1$ (corresponding to peak magnification $\gtrsim 10$), to not underestimate the expected numbers calculated below and avoid constraining $v_{\rm avg}$ too much.

An issue in the comparison is that the BH event OB110462 did not pass the selection criteria by \citet{mro17} due to its too-bright blending flux (private communication with P. Mr{\'o}z), meaning that the BH event was not ``detected" by their criteria and did not remain in the final statistical sample.
Therefore, the statement that there is at least one BH event in the sample is not true if we consider only the ``detected" events.
To include OB110462, we consider the number of events ``occurring" during the survey instead of the number of events ``detected".
This can be calculated by dividing the number of detected events, $\epsilon \, \Gamma_{\rm all}$ shown in the blue dashed curves in Fig. \ref{fig:tE_OGLEs}, by the detection efficiency of the OGLE-IV survey, $\epsilon$.
The blue solid curve in each panel of Fig. \ref{fig:tE_OGLEs} shows the $t_{\rm E}$ distribution of the number of events ``occurring" during the survey, $\Gamma_{\rm all}$.
The detection efficiency was calculated by injection and recovery tests of artificial events with $0 < u_0 < 1.5$ and $I_S < 22$ by \citet{mro17} and with $0 < u_0 < 1.0$ and $I_S < 21$ by \citet{mro19}.
Thus, $\Gamma_{\rm all}$ in the top panels of Fig. \ref{fig:tE_OGLEs} represents all events occurring with $0 < u_0 < 1.5$ and $I_S < 22$ in the OGLE-IV high-cadence fields between 2010 and 2015, while $\Gamma_{\rm all}$ in the bottom panels represents all events occurring with $0 < u_0 < 1.0$ and $I_S < 21$ in the OGLE-IV low-cadence fields between 2010 and 2017. OB110462 has $u_0 = 0.05 \pm 0.01$ \citep{lam23} and $I_S = 19.8$ mag \citep{mro22} and is supposed to be included in $\Gamma_{\rm all}$ for the high-cadence fields.

The orange solid curves in Fig. \ref{fig:tE_OGLEs} show subsamples of $\Gamma_{\rm all}$ for BH events with $u_0 < 0.1$, $\Gamma_{\rm BH} (u_0 < 0.1)$.
The sums of the expected numbers of BH events with $t_{\rm E} > 200~{\rm days}$ in both high-cadence and low-cadence fields are $\Gamma_{\rm BH} (u_0 < 0.1, t_{\rm E} > 200~{\rm days}) = $ 2.5, 2.2, 1.8, 0.88, and 0.28 for $v_{\rm avg} =$ 25 km/sec, 50 km/sec, 100 km/sec, 200 km/sec, and 400 km/sec, which are also shown in the fiducial model line in Table \ref{tab-num462} as $\Gamma_{\rm BH}$.
The $\Gamma_{\rm BH} (u_0 < 0.1, t_{\rm E} > 200~{\rm days})$ values are less than one for $v_{\rm avg} =$ 200 km/sec and 400 km/sec, suggesting that such fast $v_{\rm avg}$ might be unlikely when compared with the actual number of events occurring, at least one.
Nevertheless, this alone does not put an effective constraint on the average speed because the Poisson probabilities of having more than one event are 0.92, 0.89, 0.83, 0.59, and 0.24 for $v_{\rm avg} =$ 25 km/sec, 50 km/sec, 100 km/sec, 200 km/sec, and 400 km/sec, respectively, which all are reasonable probabilities to happen.

However, as discussed in Section \ref{sec:tEpiE}, OB110462 has a unique $\pi_{\rm E}$ value among the BH events with $t_{\rm E}$ values similar to OB110462, especially for high $v_{\rm avg}$ values (see Figure \ref{fig:tE_piEs}).
To further constrain $v_{\rm avg}$, we consider the probability of having such a rare $\pi_{\rm E}$ value when we randomly pick one BH event that has $u_0 < 0.1$ and $t_{\rm E} > 200~{\rm days}$.
Figure \ref{fig:piE462} shows the probability distribution of $\pi_{\rm E}$ for BH lens events with $259 < t_{\rm E}/{\rm days} < 282$ ($1~\sigma$ uncertainty region of $t_{\rm E}$ for OB110462) for different $v_{\rm avg}$ values.
The vertical red lines indicate the median and $3~\sigma$ range of $\pi_{\rm E}$ of OB110462 estimated by \citet{lam23}, where we take the $3~\sigma$ range ($0.06 < \pi_{\rm E} < 0.14$) from their Figure 15.
As seen in the top panel of Figure \ref{fig:piE462}, the probabilities of $\pi_{\rm E}$ larger than the $3~\sigma$ lower limit value are $P_{\rm BH} (\pi_{\rm E} > 0.06 \,|\, t_{\rm E, OB110462})= $ 0.10, 0.086, 0.054, 0.023, and $6.5 \times 10^{-3}$ for $v_{\rm avg} =$ 25 km/sec, 50 km/sec, 100 km/sec, 200 km/sec, and 400 km/sec, respectively.
When multiplying by $\Gamma_{\rm BH} (u_0 < 0.1, t_{\rm E} > 200~{\rm days})$, we have $N_{\rm exp, 0462} =$ 0.26, 0.19, 0.095, 0.020, and $1.8 \times 10^{-3}$ for $v_{\rm avg} =$ 25 km/sec, 50 km/sec, 100 km/sec, 200 km/sec, and 400 km/sec, respectively. 
Here, $N_{\rm exp, 0462}$ represents the expected number of BH events occurring with $u_0 < 0.1, t_{\rm E} > 200~{\rm days}$, and $\pi_{\rm E}$ values as rare as OB110462 during the OGLE-IV survey. Hereafter, we shortly refer to $N_{\rm exp, 0462}$ as the expected number of BH events like OB110462.

The $N_{\rm exp, 0462}$ values indicate that the average kick velocity is likely to be $v_{\rm avg} \lesssim 100$ km/sec and fast average kick velocity values like 200 km/sec or 400 km/sec are unlikely.
We would like to emphasize that our claim that such fast $v_{\rm avg}$ is unlikely is solely based on statistics using event rates, and is not directly based on a small transverse velocity of $v_{T, L} = 37.6 \pm 5.1$ km/sec \citep{lam23} for the BH lens of OB110462.

\section{Discussion} \label{sec:disc}


\subsection{Model uncertainty} \label{sec:uncertain}

Our estimate of the $N_{\rm exp, 0462}$ values in Section \ref{sec:disc_vave} may depend on the Galactic model used.
There should be no major problems with the stellar distribution in our model because the \citetalias{kos21} model is fitted to the stellar distribution toward the Galactic bulge.
However, there are some uncertainties about the parameter distributions of BHs.

The microlensing event rate is a function of number density along the line of sight, the velocity distribution, and the mass distribution of lens objects.
Among these, we have been considering various velocity distributions by changing average kick velocity values and their influences on the number density distribution of the disk BH.
However, we have so far assumed that the bulge BH profile does not change by the additional kick velocity for simplicity.
We have also fixed the BH mass distribution, which is another uncertain factor.
Here we estimate how much these uncertainties affect our estimation of $N_{\rm exp, 0462}$, the expected number of BH lens events like OB110462 during the OGLE-IV survey by \citet{mro17} and \citet{mro19}. 
The considered models and corresponding $N_{\rm exp, 0462}$ values are summarized in Table \ref{tab-num462}.


\subsubsection{Density distribution of bulge BHs} \label{sec:mod_bar}
As described in Section \ref{sec:model}, we have not been considering the influence of the kick velocity on the bulge BH density distribution because modeling it requires a dynamical simulation under the potential created by the \citetalias{kos21} model that is beyond the scope of this study.
Nevertheless, we here consider how much this uncertainty likely affects our estimation of $N_{\rm exp, 0462}$.
Fig. \ref{fig:R_hT18} shows that the additional velocity inflates the scale height even in the bulge dominant region of $R \simlt 1$ kpc.
To qualitatively see the inflation effect, we apply Eq. (\ref{eq:h_BH}) but without $g (R)$, i.e., 
\begin{align}
h_{\rm BH, b} = h_{\rm 0, b} \,\left( \frac{\sqrt{\sigma_{z, {\rm b}} (x', y', z)^2 + \pi \, v_{\rm avg}^2 /8}} {\sigma_{z, {\rm b}} (x', y', z)}\right)^{\beta_{\rm BH}}, \label{eq:h_BHb}
\end{align}
and use $h_{\rm BH, b}$ as the scale height of the bar, corresponding to $z_0$ of the \citetalias{kos21} bulge density model.
Here, $h_{\rm 0, b} = 0.24\,{\rm kpc}$ is the original scale height of the \citetalias{kos21} bulge density model and $\sigma_{z, {\rm b}} (x', y', z)$ is the velocity dispersion of the bulge stars at $(x', y', z)$
and given by Eq. (19) of \citetalias{kos21}, where $x'$ and $y'$ are the positions along the major and minor axes of the bar, respectively.

We repeat the calculation of the expected number of BH events like OB110462 and the inflate-bar model lines of Table \ref{tab-num462} show the results.
When the bulge scale height is more inflated by the kick, the event rate of bulge lens events in the direction of OB110462, i.e., $(l, b) = (-0.14, -1.62)$, becomes lower.
This leads to the smaller total event rates of BH events with $u_0 < 0.1$ and $t_{\rm E} > 200$ days, $\Gamma_{\rm BH} (u_0 < 0.1, t_{\rm E} > 200~{\rm days})$, compared to the fiducial model (shown in Table \ref{tab-num462}).
On the other hand, the decrease in bulge lenses leads to a relative increase in disk lenses, i.e., closer $D_{\rm L}$, and since $\pi_{\rm E} \propto \sqrt{D_{\rm L}^{-1}-D_{\rm S}^{-1}}$, $\pi_{\rm E}$ values also increase on average.
This results in a higher probability of $P_{\rm BH} (\pi_{\rm E} > 0.06 \,|\, t_{\rm E, OB110462})$ than the fiducial model.
The lower probability of $t_{\rm E} > 200$ days and the higher probability of $\pi_{\rm E} > 0.06$ are canceled out when multiplied, resulting in almost the same expected number of BH events like OB110462 as the fiducial model, as shown in Table \ref{tab-num462}.

Since the same logic holds whether the density of the bulge BHs increases or decreases due to the natal kick, our estimate of the expected number of BH events is robust to the ignored effect of kick velocity on the density distribution of the bulge BHs.

\subsubsection{BH mass function} \label{sec:BH_MF}
\begin{figure}
\begin{center}
\includegraphics[width=8cm]{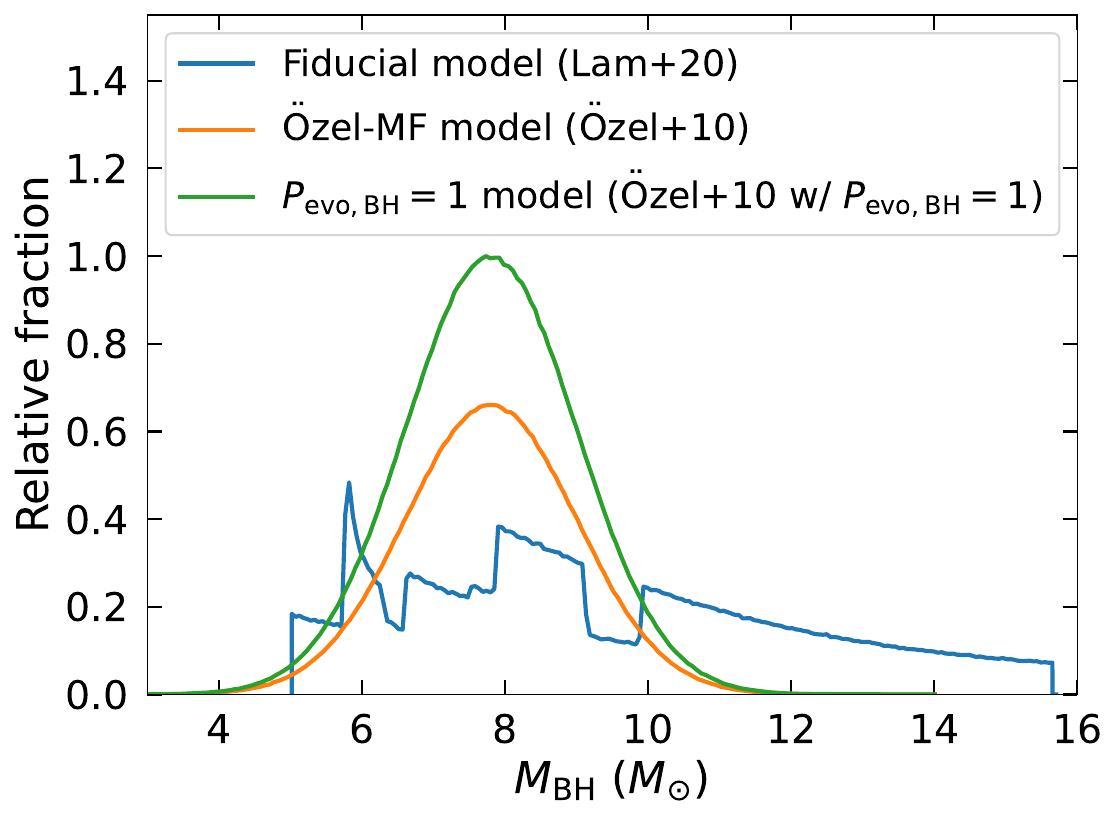}
\caption{
  \label{fig:MFBHs}
Comparison of black hole mass functions used in the three different models in Table \ref{tab-num462}. 
}
\end{center}
\end{figure}

As described in Section \ref{sec:model}, we have used the \citet{lam20}'s probabilistic initial-final mass relation (IFMR) to calculate the present-day mass function for BHs, which is shown as a blue line in Fig. \ref{fig:MFBHs}.
Another BH mass function often used is the Gaussian distribution of $7.8 \pm 1.2~M_{\odot}$ by \citet{oze10}.
The orange line in Fig. \ref{fig:MFBHs} shows the BH mass distribution when we adopt the Gaussian distribution of $7.8 \pm 1.2~M_{\odot}$ as the BH IFMR instead of the \citet{lam20}'s IFMR but holds the same probabilities of evolving into a BH, $P_{\rm evo, BH}$.
Because the latest estimated mass for the BH lens of OB110462 by \citet{lam23}, $6.0^{+1.2}_{-1.0}~M_{\odot}$, well overlaps the \citet{oze10}'s Gaussian distribution, a naive expectation is that the expected BH numbers calculated in Section \ref{sec:disc_vave} would increase if we adopt the \citet{oze10}'s BH mass function.

The {\"O}zel-MF model line in Table \ref{tab-num462} shows the recalculated expected number of BH events like OB110462.
Contrary to the naive expectation above, the event rate of BH events with $u_0 < 0.1$ and $t_{\rm E} > 200$ days, $\Gamma_{\rm BH} (u_0 < 0.1, t_{\rm E} > 200~{\rm days})$, for the {\"O}zel-MF model is lower than that for the fiducial model.
This is because the \citet{lam20}'s IFMR used in the fiducial model generates more massive BHs than the \citet{oze10}'s IFMR, and generates more events with $t_{\rm E} > 200~{\rm days}$.
Meanwhile, this in turn indicates that BH lens masses with the {\"O}zel-MF model are lower than those in the fiducial model on average, leading to higher values of $\pi_{\rm E}$ since $\pi_{\rm E} \propto M_{\rm L}^{-1/2}$.
This results in a higher probability of $P_{\rm BH} (\pi_{\rm E} > 0.06 \,|\, t_{\rm E, OB110462})$ for the {\"O}zel-MF model than the fiducial model.
Similarly to the case of the inflate-bar model in Section \ref{sec:mod_bar}, these two factors of the lower event rate and higher probability are canceled when multiplied. This again leads to almost the same expected number of BH events like OB110462 for the {\"O}zel-MF model as those for the fiducial and inflate-bar models, as shown in Table \ref{tab-num462}.
Since the same cancellation happens for any IFMR as long as the total number of BHs is the same, our estimate of the expected number of BH events is again robust to the shape of the BH mass function.

The above modification kept the fraction of stars that evolve into BHs, $P_{\rm evo, BH}$, as a function of the initial mass range that is described in Section \ref{sec:BHMFmodel}.
However, even when we apply $P_{\rm evo, BH} = 1$ for all stars with initial masses $> 15~M_{\odot}$, the number of BHs in the Galactic model only increases to 0.0054 BHs per main sequence star from the original value of 0.0035 BHs per main sequence star (see the green line in Fig. \ref{fig:MFBHs}).
Although this leads to a factor of $\sim 1.5$ increase in the expected BH events, as shown in Table \ref{tab-num462} as $P_{\rm evo, BH} = 1$ model, it does not change our conclusion that $v_{\rm avg} \geq 200~{\rm km/sec}$ is unlikely.

\subsubsection{Uncertainty in $g (R)$} \label{sec:unc_gR}
We have been using Eq. (\ref{eq:h_BH}) for the disk BH scale height model, which includes an uncertain factor $g (R)$ to connect the two extremes of the disk models; one with a constant scale height but with a velocity dispersion dependent on $R$ (when $v_{\rm avg} = 0$), and the other with a scale height dependent on $R$ but with a constant velocity dispersion (when $v_{\rm avg} \to \infty$).
How to connect these two extreme cases is uncertain, and we have been applying Eqs. (\ref{eq:gR})-(\ref{eq:RBH}).
To estimate the possible variation of the expected number of BH events due to this uncertainty, we calculate $N_{\rm exp, 0462}$ with the two extreme cases.
The $R_{\rm BH} = R_{\rm BH, 0}$ and $R_{\rm BH} \to \infty$ models in Table \ref{tab-num462} show the results, and we confirm that our conclusion does not change due to this uncertainty.

\subsubsection{Possibility of binary BH} \label{sec:unc_BinBH}
\citet{mro17} and \citet{mro19} removed obvious binary lens events from their sample and therefore the $t_{\rm E}$ distribution shown in Fig. \ref{fig:tE_OGLEs} is for events with the shape of a single lens event.
However, the sample may still contain very close binary lenses, because it is difficult to recognize a binary signal if the size of the central caustic made by the close binary is smaller than $u_0$.

Because our model has been only considering single BHs, including such unrecognized binary BHs could increase the expected number of BH events.
Assuming each component of binary BHs follows the same BH mass function as what we considered for single BHs, an equal-mass binary BH system will have its system mass function that doubles the mass of the mass function for single BHs.
Therefore, events with $t_{\rm E} \simgt 140~{\rm days}$ in a single BH-only model could potentially contribute $t_{\rm E} > 200$ days when we consider a model with binary BH population.

However, as we discussed in Section \ref{sec:BH_MF}, our estimates of $N_{\rm exp, 0462}$ are robust to any changes in the BH mass function as long as the total number of BHs is conserved. 
This is again due to the cancellation between the two factors of  $\Gamma_{\rm BH} (u_0 < 0.1, t_{\rm E} > 200~{\rm days})$ and $P_{\rm BH} (\pi_{\rm E} > 0.06 \,|\, t_{\rm E, OB110462})$ because of the different dependence on $M_{\rm L}$ for $t_{\rm E} \propto M_{\rm L}^{1/2}$ and $\pi_{\rm E} \propto M_{\rm L}^{-1/2}$.
\citet{ole20} found about 8\% of BHs are in binary systems based on a population synthesis study.
Although the uncertainty of the binary fraction is large, any reasonable amount of the BH binary fraction would not increase the total number of BHs significantly enough to change our conclusion.


\subsection{Were we lucky to have the BH event detection?} \label{sec:BH_pop}
We have so far discussed the unlikeliness of $v_{\rm avg} = 200~{\rm km/sec}$ or $v_{\rm avg} = 400~{\rm km/sec}$ due to their small values of $N_{\rm exp, 0462}$, the expected numbers of BH events like OB110462 occurring during the OGLE-IV survey by \citet{mro17} and \citet{mro19}. These are $\sim 0.02$ for $v_{\rm avg} = 200~{\rm km/sec}$ and $\sim 0.002$ for $v_{\rm avg} = 400~{\rm km/sec}$, which are much less than the actual number of such events, i.e., at least one.
On the other hand, the $N_{\rm exp, 0462}$ values for slower kick velocities are not large either. Even the value for $v_{\rm avg} = 25~{\rm km/sec}$ is 0.24 to 0.44 depending on the model, but does not exceed one. 
The interpretation of this is either of or a combination of the following.

The first possibility is that we were lucky to find the first isolated stellar-mass BH in such a unique parameter space.
This can be tested by further observations of other BH events. If we see more BH events in a similarly unlikely parameter space, it would indicate this is not the case.
This will be effectively investigated by the Nancy Grace Roman Space Telescope \citep{spe15} which will be launched by May 2027.
\citet{saj23} predicts the masses of about 20 BHs can be measured assuming the BH fraction of 0.01\% relative to the number of total objects if the Roman can fill its gap period between the first three and second three bulge seasons, even at low frequency. The number would be 400 BHs with the BH fraction of 0.2\% in our model.

The second possibility is that there were modeling issues, and we missed true, more likely, values of $t_{\rm E}$ and/or $\pi_{\rm E}$.
This seems unlikely because several independent modeling attempts have been made for OB110462 \citep{sah22, mro22, lam23}, all showing more or less similar estimates for $t_{\rm E}$ and $\pi_{\rm E}$ values.
Nevertheless, it might be worth trying to include higher-order effects that have not been reported in the papers, such as a binary lens or the xallarap effect, to see how much the estimates of $t_{\rm E}$ or $\pi_{\rm E}$ would change.

The third possibility is that there are unconsidered issues in our calculation of $N_{\rm exp, 0462}$.
Although we have reviewed likely uncertainties in our model in Section \ref{sec:uncertain}, there could be uncertainties in parts where we assumed robustness.
One possibility is an overestimation of the detection efficiency of the OGLE-IV survey for long-timescale single-lens events by \citet{mro17} and \citet{mro19}, which was also pointed out by \citet{mro19}.
Although the microlens parallax effect is usually very subtle and does not affect the detectability of single-lens events, it could sometimes cause a significant deviation from the Paczy{\'n}ski curve particularly for long-timescale events with $t_{\rm E} \simgt 100~{\rm days}$ \citep[some examples are seen in][]{wyr16}.
Meanwhile, Mr{\'o}z et al. estimated the detection efficiency by embedding many artificial single-lens events and counting the number of events passing their selection criteria. Either the artificial events or the single-lens model used in the selection criteria does not include the parallax effect. 
This results in an overestimation of the detection efficiency for long-timescale events whose detectability could be affected by the parallax effect because the real events always have non-zero $\pi_{\rm E}$ values.
The overestimation of the detection efficiency causes an underestimation of $N_{\rm exp, 0462}$, because our $\Gamma_{\rm BH} (u_0 < 0.1, t_{\rm E} > 200~{\rm days})$ values have been calculated by dividing the expected value of event detection that can reproduce the OGLE-IV $t_{\rm E}$ distribution by the detection efficiency.
Although the presence of an overestimation of the detection efficiency is plausible, the relative values of $N_{\rm exp, 0462}$ among different $v_{\rm avg}$ values still hold, and it is unlikely that correcting this overestimation alone would increase the $N_{\rm exp, 0462}$ values by more than one order of magnitude.
Therefore, our conclusion that $v_{\rm avg} = 200~{\rm km/sec}$ or $v_{\rm avg} = 400~{\rm km/sec}$ is unlikely does not change due to this, whereas the $N_{\rm exp, 0462}$ values smaller than 1 even for $v_{\rm avg} = 25~{\rm km/sec}$ might be mitigated by this.

\subsection{Comparison with previous studies} 
There are some attempts to constrain the kick velocity of compact objects from the observations of microlensing events caused by them.  For example, \cite{2022ApJ...930..159A} gave a constraint on the velocity of the natal kick to a BH that caused OB110462 from its inferred peculiar velocity as $\lesssim 100~{\rm km}~{\rm s}^{-1}$.  Although their conclusion is similar to ours, there is a difference that our analysis is not based on the peculiar velocity of the lens but based on statistical arguments about the event rate with observed $t_{\rm E}$ and $\pi_{\rm E}$.  The impact of the natal kick to compact remnants on the observations of photometric microlensing was partially investigated by \cite{2021AcA....71...89M}.  However, they assumed that all the remnants receive the same kick velocity and do not consider the effect of spatial distribution changes due to kicks.  Recently, \cite{2024arXiv240314612S} made predictions regarding gravitational microlensing phenomena caused by compact remnants taking into account changes in spatial distribution and velocity distribution in the Galaxy due to kicks.  However, they conducted the analysis based on a single kick model, thus not being able to constrain kick velocities from observations.  We predict the duration ($t_{\rm E}$) and parallax ($\pi_{\rm E}$) distribution of gravitational microlensing caused by BHs considering various average kick velocities, while incorporating the spatial distribution and velocity distribution of them.  The influence of the kick velocity on the spatial distribution has not been considered in previous population synthesis models for microlensing such as the ones by \citet{lam20} or \citetalias{kos21}.  Furthermore, we discuss the constraints on the kick velocity and the origin of a BH by comparing our results with the observations of OB110462. The results obtained in this study can also be applied in future observations when a large number of gravitational microlensing events are detected with the Roman telescope, for example.

\section{Summary}
In this paper, we have investigated how the natal kick velocity for Galactic BHs affects the microlensing event rate distribution for the BHs.
We considered a Maxwell distribution for the kick velocity with averages of $v_{\rm avg} =$ 25 km/sec, 50 km/sec, 100 km/sec, 200 km/sec, and 400 km/sec, as well as the consequent inflation of the scale height and variation of the surface density of the BH disk.
We found that the event rate for the disk BH lenses toward the Galactic bulge decreases as $v_{\rm avg}$ increases, mainly due to the velocity-dependent scale height inflation which has not been taken into account by previous studies. 

We then focused on OB110462, the only unambiguous isolated stellar-mass BH event to date.
We calculated the expected number of BH events with parameters similar to OB110462 occurring during the OGLE-IV survey by \citet{mro17} and \citet{mro19}, which is likely a larger sample than the one searched by a series of HST programs by \citet{sah09} that discovered OB110462.
Our fiducial model predicted the expected number of BH events like OB110462 as $N_{\rm exp, 0462} =$ 0.26, 0.19, 0.095, 0.020, and $1.8 \times 10^{-3}$ for $v_{\rm avg} =$ 25 km/sec, 50 km/sec, 100 km/sec, 200 km/sec, and 400 km/sec, respectively, and these numbers were not sensitive to the BH mass function or the choice of models as far as we considered.
The $N_{\rm exp, 0462}$ values could be underestimated because we did not consider the likely overestimation of the detection efficiency of the OGLE-IV survey for long-timescale events \citep{mro19}.
However, it is unlikely that the numbers would be more than 10 times larger even if the overestimation of the detection efficiency is corrected.
Therefore, we concluded that the average kick velocity is likely to be $v_{\rm avg} \lesssim 100~{\rm km/sec}$.

The $N_{\rm exp, 0462}$ value takes its maximum of 0.26 for the lowest kick of $v_{\rm avg} =$ 25 km/sec, which still expects less than one event.
This might indicate that we were lucky to discover the first BH event in such a unique parameter space, but another possibility is due to the underestimation of $N_{\rm exp, 0462}$ because of the overestimation of the detection efficiency.
The possibility of this event as a statistical fluke can be confirmed/confuted by whether or not we see more BH events in a similarly unlikely parameter space in the future.

The microlensing survey by the Nancy Grace Roman Space Telescope \citep{spe15} is capable of measuring masses of 20 BHs \citep{saj23} when the BH fraction in our Galaxy is assumed to be 0.01\%, and this would be 400 BHs when a 0.2\% BH fraction is assumed as in our model.
We can extend this study using the BH samples by the Roman survey to constrain the kick velocity distribution even tighter.

\acknowledgments
The work of N. Koshimoto was supported by the JSPS overseas research fellowship and JSPS KAKENHI Grant Numbers JP24K17089 and JP23KK0060. The work of N. Kawanaka was supported by the JSPS KAKENHI Grant Numbers JP22K03686. DT is supported by the Sherman Fairchild Postdoctoral Fellowship at California Institute of Technology.  We thank Przemek Mr{\'o}z for providing the detection efficiency data and other useful information. We thank Ataru Tanikawa and Junichi Baba for their useful comments. 


\begin{thebibliography}{}

\bibitem[Agol \& Kamionkowski(2002)]{2002MNRAS.334..553A} Agol, E. \& Kamionkowski, M.\ 2002, \mnras, 334, 553. doi:10.1046/j.1365-8711.2002.05523.x

\bibitem[Agol et al.(2002)]{2002ApJ...576L.131A} Agol, E., Kamionkowski, M., Koopmans, L.~V.~E., et al.\ 2002, \apjl, 576, L131. doi:10.1086/343758

\bibitem[Andrews et al.(2019)]{2019ApJ...886...68A} Andrews, J.~J., Breivik, K., \& Chatterjee, S.\ 2019, \apj, 886, 68. doi:10.3847/1538-4357/ab441f

\bibitem[Andrews \& Kalogera(2022)]{2022ApJ...930..159A} Andrews, J.~J. \& Kalogera, V.\ 2022, \apj, 930, 159. doi:10.3847/1538-4357/ac66d6

\bibitem[Andrews et al.(2022)]{2022arXiv220700680A} Andrews, J.~J., Taggart, K., \& Foley, R.\ 2022, arXiv:2207.00680. doi:10.48550/arXiv.2207.00680

\bibitem[Armitage \& Natarajan(1999)]{1999ApJ...523L...7A} Armitage, P.~J. \& Natarajan, P.\ 1999, \apjl, 523, L7. doi:10.1086/312261

\bibitem[Bailyn et al.(1998)]{1998ApJ...499..367B} Bailyn, C.~D., Jain, R.~K., Coppi, P., et al.\ 1998, \apj, 499, 367. doi:10.1086/305614

\bibitem[Banagiri et al.(2023)]{2023ApJ...959..106B} Banagiri, S., Doctor, Z., Kalogera, V., et al.\ 2023, \apj, 959, 106. doi:10.3847/1538-4357/ad0557

\bibitem[Barkov et al.(2012)]{2012MNRAS.427..589B} Barkov, M.~V., Khangulyan, D.~V., \& Popov, S.~B.\ 2012, \mnras, 427, 589. doi:10.1111/j.1365-2966.2012.22029.x

\bibitem[Bennett et al.(2002)]{ben02} Bennett, D.~P., Becker, A.~C., Quinn, J.~L., et al.\ 2002, \apj, 579, 639. doi:10.1086/342225

\bibitem[Blaineau et al.(2022)]{bla22} Blaineau, T., Moniez, M., Afonso, C., et al.\ 2022, \aap, 664, A106. doi:10.1051/0004-6361/202243430

\bibitem[Bovy(2017)]{bov17} Bovy, J.\ 2017, \mnras, 470, 1360. doi:10.1093/mnras/stx1277

\bibitem[Breivik et al.(2017)]{2017ApJ...850L..13B} Breivik, K., Chatterjee, S., \& Larson, S.~L.\ 2017, \apjl, 850, L13. doi:10.3847/2041-8213/aa97d5

\bibitem[Campana \& Pardi(1993)]{1993A&A...277..477C} Campana, S. \& Pardi, M.~C.\ 1993, \aap, 277, 477

\bibitem[Caputo et al.(2017)]{2017MNRAS.468.4000C}
Caputo, D.~P., de Vries, N., Patruno, A., et al.\ 2017, \mnras, 468, 4000. doi:10.1093/mnras/stw3336

\bibitem[Carr(1979)]{1979MNRAS.189..123C} Carr, B.~J.\ 1979, \mnras, 189, 123. doi:10.1093/mnras/189.1.123

\bibitem[Chawla et al.(2022)]{2022ApJ...931..107C} Chawla, C., Chatterjee, S., Breivik, K., et al.\ 2022, \apj, 931, 107. doi:10.3847/1538-4357/ac60a5

\bibitem[Chisholm et al.(2003)]{2003ApJ...596..437C} Chisholm, J.~R., Dodelson, S., \& Kolb, E.~W.\ 2003, \apj, 596, 437. doi:10.1086/377628

\bibitem[Chung et al.(2005)]{chu05} Chung, S.-J., Han, C., Park, B.-G., et al.\ 2005, \apj, 630, 535. doi:10.1086/432048

\bibitem[Clarke et al.(2019)]{cla19} Clarke, J.~P., Wegg, C., Gerhard, O., et al.\ 2019, \mnras, 489, 3519. doi:10.1093/mnras/stz2382

\bibitem[Corral-Santana et al.(2016)]{2016A&A...587A..61C} Corral-Santana, J.~M., Casares, J., Mu{\~n}oz-Darias, T., et al.\ 2016, \aap, 587, A61. doi:10.1051/0004-6361/201527130

\bibitem[Einstein(1936)]{1936Sci....84..506E} Einstein, A.\ 1936, Science, 84, 506. doi:10.1126/science.84.2188.506

\bibitem[El-Badry et al.(2023a)]{2023MNRAS.518.1057E} El-Badry, K., Rix, H.-W., Quataert, E., et al.\ 2023, \mnras, 518, 1057. doi:10.1093/mnras/stac3140

\bibitem[El-Badry et al.(2023b)]{2023MNRAS.521.4323E} El-Badry, K., Rix, H.-W., Cendes, Y., et al.\ 2023, \mnras, 521, 4323. doi:10.1093/mnras/stad799

\bibitem[Farr et al.(2011)]{2011ApJ...741..103F} Farr, W.~M., Sravan, N., Cantrell, A., et al.\ 2011, \apj, 741, 103. doi:10.1088/0004-637X/741/2/103

\bibitem[Fender et al.(2013)]{2013MNRAS.430.1538F} Fender, R.~P., Maccarone, T.~J., \& Heywood, I.\ 2013, \mnras, 430, 1538. doi:10.1093/mnras/sts688

\bibitem[Fragos et al.(2009)]{2009ApJ...697.1057F} Fragos, T., Willems, B., Kalogera, V., et al.\ 2009, \apj, 697, 1057. doi:10.1088/0004-637X/697/2/1057

\bibitem[Fryer et al.(2012)]{2012ApJ...749...91F} Fryer, C.~L., Belczynski, K., Wiktorowicz, G., et al.\ 2012, \apj, 749, 91. doi:10.1088/0004-637X/749/1/91

\bibitem[Fujita et al.(1998)]{1998ApJ...495L..85F} Fujita, Y., Inoue, S., Nakamura, T., et al.\ 1998, \apjl, 495, L85. doi:10.1086/311220

\bibitem[Gaia Collaboration et al.(2018)]{kat18} Gaia Collaboration, Katz, D., Antoja, T., et al.\ 2018, \aap, 616, A11

\bibitem[Gaia Collaboration et al.(2024)]{2024arXiv240410486G} Gaia Collaboration, Panuzzo, P., Mazeh, T., et al.\ 2024, arXiv:2404.10486

\bibitem[Gandhi et al.(2020)]{2020MNRAS.496L..22G} Gandhi, P., Rao, A., Charles, P.~A., et al.\ 2020, \mnras, 496, L22. doi:10.1093/mnrasl/slaa081

\bibitem[Gould(2000)]{2000ApJ...535..928G} Gould, A.\ 2000, \apj, 535, 928. doi:10.1086/308865

\bibitem[Grindlay(1978)]{1978ApJ...221..234G} Grindlay, J.~E.\ 1978, \apj, 221, 234. doi:10.1086/156023

\bibitem[Hobbs et al.(2005)]{2005MNRAS.360..974H} Hobbs, G., Lorimer, D.~R., Lyne, A.~G., et al.\ 2005, \mnras, 360, 974. doi:10.1111/j.1365-2966.2005.09087.x

\bibitem[Hog et al.(1995)]{1995A&A...294..287H} Hog, E., Novikov, I.~D., \& Polnarev, A.~G.\ 1995, \aap, 294, 287

\bibitem[Igoshev(2020)]{2020MNRAS.494.3663I} Igoshev, A.~P.\ 2020, \mnras, 494, 3663. doi:10.1093/mnras/staa958

\bibitem[Igoshev et al.(2021)]{2021MNRAS.508.3345I} Igoshev, A.~P., Chruslinska, M., Dorozsmai, A., et al.\ 2021, \mnras, 508, 3345. doi:10.1093/mnras/stab2734

\bibitem[Ioka et al.(2017)]{2017MNRAS.470.3332I} Ioka, K., Matsumoto, T., Teraki, Y., et al.\ 2017, \mnras, 470, 3332. doi:10.1093/mnras/stx1337

\bibitem[Jonker et al.(2021)]{2021ApJ...921..131J} Jonker, P.~G., Kaur, K., Stone, N., et al.\ 2021, \apj, 921, 131. doi:10.3847/1538-4357/ac2839

\bibitem[Kains et al.(2017)]{2017ApJ...843..145K}
Kains, N., Calamida, A., Sahu, K.~C., et al.\ 2017, \apj, 843, 145. doi:10.3847/1538-4357/aa78eb

\bibitem[Kawanaka et al.(2017)]{2017IAUS..324...41K} Kawanaka, N., Yamaguchi, M., Piran, T., et al.\ 2017, New Frontiers in Black Hole Astrophysics, 324, 41. doi:10.1017/S1743921316012606

\bibitem[Kimball et al.(2023)]{2023ApJ...952L..34K} Kimball, C., Imperato, S., Kalogera, V., et al.\ 2023, \apjl, 952, L34. doi:10.3847/2041-8213/ace526

\bibitem[Kimura et al.(2021)]{2021ApJ...922L..15K} Kimura, S.~S., Kashiyama, K., \& Hotokezaka, K.\ 2021, \apjl, 922, L15. doi:10.3847/2041-8213/ac35dc

\bibitem[Kinugawa \& Yamaguchi(2018)]{2018arXiv181009721K} Kinugawa, T. \& Yamaguchi, M.~S.\ 2018, arXiv:1810.09721. doi:10.48550/arXiv.1810.09721

\bibitem[Koshimoto et al.(2020)]{kos20} Koshimoto, N., Bennett, D.~P., \& Suzuki, D.\ 2020, \aj, 159, 268. doi:10.3847/1538-3881/ab8adf

\bibitem[Koshimoto et al. (2021)]{kos21} Koshimoto, N., Baba, J., \& Bennett, D.~P.\ 2021, \apj, 917, 78. doi:10.3847/1538-4357/ac07a8 \citepalias{kos21}

\bibitem[Koshimoto \& Ranc(2022)]{kosran22} Koshimoto, N. \& Ranc, C.\ 2022, Zenodo.4784948. doi:10.5281/zenodo.4784948

\bibitem[Kunder et al.(2012)]{kun12} Kunder, A., Koch, A., Rich, R.~M., et al.\ 2012, \aj, 143, 57. doi:10.1088/0004-6256/143/3/57

\bibitem[Lai et al.(2001)]{2001ApJ...549.1111L} Lai, D., Chernoff, D.~F., \& Cordes, J.~M.\ 2001, \apj, 549, 1111. doi:10.1086/319455

\bibitem[Lam \& Lu(2023)]{lam23} Lam, C.~Y. \& Lu, J.~R.\ 2023, arXiv:2308.03302. doi:10.48550/arXiv.2308.03302

\bibitem[Lam et al.(2020)]{lam20} Lam, C.~Y., Lu, J.~R., Hosek, M.~W., et al.\ 2020, \apj, 889, 31. doi:10.3847/1538-4357/ab5fd3

\bibitem[Lam et al.(2022a)]{lam22} Lam, C.~Y., Lu, J.~R., Udalski, A., et al.\ 2022, \apjl, 933, L23. doi:10.3847/2041-8213/ac7442

\bibitem[Lam et al.(2022b)]{lam22S} Lam, C.~Y., Lu, J.~R., Udalski, A., et al.\ 2022, \apjs, 260, 55. doi:10.3847/1538-4365/ac7441

\bibitem[Lamberts et al.(2018)]{2018MNRAS.480.2704L}
Lamberts, A., Garrison-Kimmel, S., Hopkins, P.~F., et al.\ 2018, \mnras, 480, 2704. doi:10.1093/mnras/sty2035

\bibitem[Lu et al.(2016)]{2016ApJ...830...41L}
Lu, J.~R., Sinukoff, E., Ofek, E.~O., et al.\ 2016, \apj, 830, 41. doi:10.3847/0004-637X/830/1/41

\bibitem[Maccarone(2005)]{2005MNRAS.360L..30M} Maccarone, T.~J.\ 2005, \mnras, 360, L30. doi:10.1111/j.1745-3933.2005.00039.x

\bibitem[Mashian \& Loeb(2017)]{2017MNRAS.470.2611M}
Mashian, N. \& Loeb, A.\ 2017, \mnras, 470, 2611. doi:10.1093/mnras/stx1410

\bibitem[Matsumoto et al.(2018)]{2018MNRAS.475.1251M} Matsumoto, T., Teraki, Y., \& Ioka, K.\ 2018, \mnras, 475, 1251. doi:10.1093/mnras/stx3148

\bibitem[McDowell(1985)]{1985MNRAS.217...77M} McDowell, J.\ 1985, \mnras, 217, 77. doi:10.1093/mnras/217.1.77

\bibitem[Meszaros(1975)]{1975A&A....44...59M} Meszaros, P.\ 1975, \aap, 44, 59

\bibitem[Mii \& Totani(2005)]{2005ApJ...628..873M} Mii, H. \& Totani, T.\ 2005, \apj, 628, 873. doi:10.1086/430942

\bibitem[Miyamoto \& Yoshii(1995)]{1995AJ....110.1427M} Miyamoto, M. \& Yoshii, Y.\ 1995, \aj, 110, 1427. doi:10.1086/117616

\bibitem[Mr{\'o}z et al.(2017)]{mro17} Mr{\'o}z, P., Udalski, A., Skowron, J., et al.\ 2017, \nat, 548, 183 

\bibitem[Mr{\'o}z et al.(2019)]{mro19} Mr{\'o}z, P., Udalski, A., Skowron, J., et al.\ 2019, \apjs, 244, 29. doi:10.3847/1538-4365/ab426b

\bibitem[Mr{\'o}z et al.(2022)]{mro22} Mr{\'o}z, P., Udalski, A., \& Gould, A.\ 2022, \apjl, 937, L24. doi:10.3847/2041-8213/ac90bb

\bibitem[Mroz et al.(2021)]{2021arXiv210713697M} Mroz, P., Udalski, A., Wyrzykowski, L., et al.\ 2021, arXiv:2107.13697. doi:10.48550/arXiv.2107.13697

\bibitem[Mr{\'o}z \& Wyrzykowski(2021)]{2021AcA....71...89M} Mr{\'o}z, P. \& Wyrzykowski, {\L}.\ 2021, \actaa, 71, 89. doi:10.32023/0001-5237/71.2.1

\bibitem[Nataf et al.(2013)]{nat13} Nataf, D.~M., Gould, A., Fouqu{\'e}, P., et al.\ 2013, \apj, 769, 88. doi:10.1088/0004-637X/769/2/88

\bibitem[Niikura et al.(2019)]{nii19} Niikura, H., Takada, M., Yokoyama, S., et al.\ 2019, \prd, 99, 083503. doi:10.1103/PhysRevD.99.083503

\bibitem[Olejak et al.(2020)]{ole20} Olejak, A., Belczynski, K., Bulik, T., et al.\ 2020, \aap, 638, A94. doi:10.1051/0004-6361/201936557

\bibitem[{\"O}zel et al.(2010)]{oze10}
{\"O}zel, F., Psaltis, D., Narayan, R., et al.\ 2010, \apj, 725, 1918. doi:10.1088/0004-637X/725/2/1918

\bibitem[Paczynski(1986)]{1986ApJ...304....1P}
Paczynski, B.\ 1986, \apj, 304, 1. doi:10.1086/164140

\bibitem[Perkins et al.(2024)]{2024ApJ...961..179P} Perkins, S.~E., McGill, P., Dawson, W., et al.\ 2024, \apj, 961, 179. doi:10.3847/1538-4357/ad09bf

\bibitem[Popov \& Prokhorov(1998)]{1998A&A...331..535P} Popov, S.~B. \& Prokhorov, M.~E.\ 1998, \aap, 331, 535. doi:10.48550/arXiv.astro-ph/9705236

\bibitem[Raithel et al.(2018)]{rai18} Raithel, C.~A., Sukhbold, T., \& {\"O}zel, F.\ 2018, \apj, 856, 35. doi:10.3847/1538-4357/aab09b

\bibitem[Rich et al.(2007)]{ric07} Rich, R.~M., Reitzel, D.~B., Howard, C.~D., et al.\ 2007, \apjl, 658, L29. doi:10.1086/513509

\bibitem[Rose et al.(2022)]{ros22} Rose, S., Lam, C.~Y., Lu, J.~R., et al.\ 2022, \apj, 941, 116. doi:10.3847/1538-4357/aca09d

\bibitem[Sahu(2009)]{sah09} Sahu, K.\ 2009, HST Proposal, 11707

\bibitem[Sahu et al.(2017)]{2017Sci...356.1046S}
Sahu, K.~C., Anderson, J., Casertano, S., et al.\ 2017, Science, 356, 1046. doi:10.1126/science.aal2879

\bibitem[Sahu et al.(2022)]{sah22} Sahu, K.~C., Anderson, J., Casertano, S., et al.\ 2022, \apj, 933, 83. doi:10.3847/1538-4357/ac739e

\bibitem[Sajadian \& Sahu(2023)]{saj23} Sajadian, S. \& Sahu, K.~C.\ 2023, \aj, 165, 96. doi:10.3847/1538-3881/acb20f

\bibitem[Samland(1998)]{1998ApJ...496..155S}
Samland, M.\ 1998, \apj, 496, 155. doi:10.1086/305368

\bibitem[Sartore \& Treves(2010)]{2010A&A...523A..33S} Sartore, N. \& Treves, A.\ 2010, \aap, 523, A33. doi:10.1051/0004-6361/201015060

\bibitem[Shahaf et al.(2023)]{2023MNRAS.518.2991S} Shahaf, S., Bashi, D., Mazeh, T., et al.\ 2023, \mnras, 518, 2991. doi:10.1093/mnras/stac3290

\bibitem[Shao \& Li(2019)]{2019ApJ...885..151S} Shao, Y. \& Li, X.-D.\ 2019, \apj, 885, 151. doi:10.3847/1538-4357/ab4816

\bibitem[Shapiro \& Teukolsky(1983)]{1983bhwd.book.....S}
Shapiro, S.~L. \& Teukolsky, S.~A.\ 1983, A Wiley-Interscience Publication, New York: Wiley, 1983

\bibitem[Shikauchi et al.(2020)]{2020PASJ...72...45S} Shikauchi, M., Kumamoto, J., Tanikawa, A., et al.\ 2020, \pasj, 72, 45. doi:10.1093/pasj/psaa030

\bibitem[Shikauchi et al.(2022)]{shikauchi+22}
Shikauchi, M., Tanikawa, A. \& Kawanaka, N.\ 2022, \apj, 928, 13

\bibitem[Shikauchi et al.(2023)]{shikauchi+23}
Shikauchi, M., Tsuna, D., Tanikawa, A. \& Kawanaka, N.\ 2023, \apj, 953, 52

\bibitem[Shvartsman(1971)]{1971SvA....15..377S} Shvartsman, V.~F.\ 1971, \sovast, 15, 377

\bibitem[Smith et al.(2018)]{smi18} Smith, L.~C., Lucas, P.~W., Kurtev, R., et al.\ 2018, \mnras, 474, 1826. doi:10.1093/mnras/stx2789

\bibitem[Spergel et al.(2015)]{spe15} Spergel, D., Gehrels, N., Baltay, C., et al.\ 2015, arXiv:1503.03757. doi:10.48550/arXiv.1503.03757

\bibitem[Sukhbold et al.(2016)]{suk16} Sukhbold, T., Ertl, T., Woosley, S.~E., et al.\ 2016, \apj, 821, 38. doi:10.3847/0004-637X/821/1/38

\bibitem[Sweeney et al.(2024)]{2024arXiv240314612S} Sweeney, D., Tuthill, P., Krone-Martins, A., et al.\ 2024, arXiv:2403.14612. doi:10.48550/arXiv.2403.14612

\bibitem[Sweeney et al.(2022)]{2022MNRAS.516.4971S} Sweeney, D., Tuthill, P., Sharma, S., et al.\ 2022, \mnras, 516, 4971. doi:10.1093/mnras/stac2092

\bibitem[Tanikawa et al.(2023)]{2023ApJ...946...79T} Tanikawa, A., Hattori, K., Kawanaka, N., et al.\ 2023, \apj, 946, 79. doi:10.3847/1538-4357/acbf36

\bibitem[Tsuna et al.(2018)]{tsu18} Tsuna, D., Kawanaka, N., \& Totani, T.\ 2018, \mnras, 477, 791. doi:10.1093/mnras/sty699

\bibitem[Tsuna \& Kawanaka(2019)]{2019MNRAS.488.2099T} Tsuna, D. \& Kawanaka, N.\ 2019, \mnras, 488, 2099. doi:10.1093/mnras/stz1809

\bibitem[van den Heuvel(1992)]{1992eocm.rept...29V}
van den Heuvel, E.~P.~J.\ 1992, In ESA, Environment Observation and Climate Modelling Through International Space Projects. Space Sciences with Particular Emphasis on High-Energy Astrophysics p 29-36 (SEE N93-23878 08-88)

\bibitem[van der Kruit \& Freeman(2011)]{van11} van der Kruit, P.~C. \& Freeman, K.~C.\ 2011, \araa, 49, 301. doi:10.1146/annurev-astro-083109-153241

\bibitem[Vigna-G{\'o}mez \& Ramirez-Ruiz(2023)]{2023ApJ...946L...2V} Vigna-G{\'o}mez, A. \& Ramirez-Ruiz, E.\ 2023, \apjl, 946, L2. doi:10.3847/2041-8213/acc076

\bibitem[Walker(1995)]{1995ApJ...453...37W} Walker, M.~A.\ 1995, \apj, 453, 37. doi:10.1086/176367

\bibitem[Wiktorowicz et al.(2020)]{2020ApJ...905..134W} Wiktorowicz, G., Lu, Y., Wyrzykowski, {\L}., et al.\ 2020, \apj, 905, 134. doi:10.3847/1538-4357/abc699

\bibitem[Willems et al.(2005)]{2005ApJ...625..324W} Willems, B., Henninger, M., Levin, T., et al.\ 2005, \apj, 625, 324. doi:10.1086/429557

\bibitem[Wong et al.(2014)]{2014ApJ...790..119W} Wong, T.-W., Valsecchi, F., Ansari, A., et al.\ 2014, \apj, 790, 119. doi:10.1088/0004-637X/790/2/119

\bibitem[Wong et al.(2012)]{2012ApJ...747..111W} Wong, T.-W., Valsecchi, F., Fragos, T., et al.\ 2012, \apj, 747, 111. doi:10.1088/0004-637X/747/2/111

\bibitem[Wyrzykowski \& Mandel(2020)]{2020A&A...636A..20W} Wyrzykowski, {\L}. \& Mandel, I.\ 2020, \aap, 636, A20. doi:10.1051/0004-6361/201935842

\bibitem[Wyrzykowski et al.(2016)]{wyr16} Wyrzykowski, {\L}., Kostrzewa-Rutkowska, Z., Skowron, J., et al.\ 2016, \mnras, 458, 3012. doi:10.1093/mnras/stw426

\bibitem[Yalinewich et al.(2018)]{2018MNRAS.481..930Y}
Yalinewich, A., Beniamini, P., Hotokezaka, K., et al.\ 2018, \mnras, 481, 930. doi:10.1093/mnras/sty2327

\bibitem[Yamaguchi et al.(2018)]{2018ApJ...861...21Y}
Yamaguchi, M.~S., Kawanaka, N., Bulik, T., et al.\ 2018, \apj, 861, 21. doi:10.3847/1538-4357/aac5ec

\bibitem[Zhao et al.(2023)]{2023MNRAS.525.1498Z} Zhao, Y., Gandhi, P., Dashwood Brown, C., et al.\ 2023, \mnras, 525, 1498. doi:10.1093/mnras/stad2226

\end{thebibliography}
\end{document}